\documentclass[]{aa}
\usepackage[varg]{txfonts}
\usepackage{graphicx}
\begin{document}

\title{Stellar masses and disk properties of Lupus young stellar objects traced by velocity-aligned stacked ALMA $^{13}$CO and C$^{18}$O spectra}
\author{Hsi-Wei Yen\inst{\ref{inst1}}, Patrick M. Koch\inst{\ref{inst2}}, Carlo F. Manara\inst{\ref{inst1}}, Anna Miotello\inst{\ref{inst1},\ref{inst3}}, Leonardo Testi\inst{\ref{inst1},\ref{inst4},\ref{inst5}}}

\institute{European Southern Observatory (ESO), Karl-Schwarzschild-Str. 2, D-85748 Garching, Germany; \email{hyen@eso.org}\label{inst1}
\and Academia Sinica Institute of Astronomy and Astrophysics, P.O. Box 23-141, Taipei 10617, Taiwan\label{inst2}
\and Leiden Observatory, Leiden University, Niels Bohrweg 2, NL-2333 CA Leiden, The Netherlands\label{inst3}
\and  INAF/Osservatorio Astrofisico di Arcetri, Largo Enrico Fermi 5, I-50125 Florence, Italy\label{inst4}
\and Excellence Cluster `Universe', Boltzmann-Str. 2, D-85748 Garching, Germany\label{inst5}
}
\date{Received date / Accepted date}

\abstract
{}
{Large samples of protoplanetary disks have been observed in recent ALMA surveys. The gas distributions and velocity structures of most of the disks can still not be imaged at high S/N ratios because of the short integration time per source in these surveys. In this work, we apply the velocity-aligned stacking method to extract more information from molecular-line data of these ALMA surveys and to study the kinematics and disk properties traced by molecular lines.} 
{We re-analyzed the ALMA $^{13}$CO (3--2) and C$^{18}$O (3--2) data of 88 young stellar objects (YSOs) in Lupus with the velocity-aligned stacking method. This method aligns spectra at different positions in a disk based on the projected Keplerian velocities at their positions and then stacks them. This method enhances the S/N ratios of molecular-line data and allows us to obtain better detections and to constrain dynamical stellar masses and disk orientations. 
}
{We obtain $^{13}$CO detections in 41 disks and C$^{18}$O detections in 18 disks with 11 new detections in $^{13}$CO and 9 new detections in C$^{18}$O after applying the method. We estimate the disk orientations and the dynamical masses of the central YSOs from the $^{13}$CO data. 
Our estimated dynamical stellar masses correlate with the spectroscopic stellar masses, and in a subsample of 16 sources, where the inclination angles are better constrained, the two masses are in a good agreement within the uncertainties and with a mean difference of 0.15 $M_\sun$. With more detections of fainter disks, 
our results show that high gas masses derived from the $^{13}$CO and C$^{18}$O lines tend to be associated with high dust masses estimated from the continuum emission. Nevertheless, the scatter is large and is estimated to be 0.9 dex, implying large uncertainties in deriving the disk gas mass from the line fluxes. We find that with such large uncertainties it is expected that there is no correlation between the disk gas mass and the mass accretion rate with the current data. Deeper observations to detect disks with gas masses $<$10$^{-5}$ $M_\sun$ in molecular lines are needed to investigate the correlation between the disk gas mass and the mass accretion rate.}
{}


\titlerunning{Stellar Masses and Disk Properties of Lupus YSOs}
\authorrunning{H.-W. Yen et al.}

\maketitle

\section{Introduction}
Protoplanetary disks are considered to be the sites of planet formation \citep[e.g.,][]{Williams11}. 
Statistical studies on properties and evolution of protoplanetary disks are essential to shed light on the diverse properties of exoplanetary systems \citep[e.g.,][]{Winn15}.
With the unprecedented sensitivity of the Atacama Large Millimeter/submillimeter Array (ALMA), 
several tens of protoplanetary disks can be imaged in continuum at sub-arcsecond resolutions and at high signal-to-noise (S/N) ratios in a few hours, providing large samples for statistical studies. 
Such ALMA surveys have been conducted toward a few star-forming regions, including Upper Scorpius \citep{Carpenter14,Barenfeld16}, Lupus \citep{Ansdell16}, Chamaeleon I \citep{Pascucci16}, and $\sigma$ Orionis \citep{Ansdell17}. 
With these large samples, the relations between the disk dust mass and the stellar mass in the individual star-forming regions have been revealed, 
and there is a trend that the relation becomes steeper with age. 
This can be explained with grain growth, drift, and fragmentation \citep{Pascucci16, Ansdell17}.
Nevertheless, the disk dust mass estimated from (sub-)millimeter continuum emission can be underestimated by a factor of a few due to uncertainties in the optical depth of the continuum emission and the maximum grain size \citep{Dunham14, Tsukamoto17}.
The high-resolution and high-sensitivity continuum data also allow detailed analyses of physical structures of disks,   
and disks in older star-forming regions tend to be less massive and larger than those in younger star-forming regions \citep{Tazzari17}. 
In addition, in synergy with the ALMA and VLT/X-shooter surveys, 
the disk dust mass and the mass accretion rate estimated from the X-shooter spectra in the Lupus and Chamaeleon~I regions are found to be correlated \citep{Manara16, Mulders17}. 
The overall properties of dusty disk populations appear to be consistent with some general features predicted by viscous accretion disk models under the condition that the viscous timescale is of the order of $\sim$1 Myr \citep{Lodato17, Mulders17, Rosotti17}. 
Detailed investigations of the gaseous disk properties are essential to verify these initial findings.

To investigate gas components in disks, CO isotopologue lines are also observed simultaneously in these ALMA surveys,   
but less than one third of the disks are detected in the CO isotopologue lines \citep{Ansdell16, Ansdell17, Long17}. 
 The total fluxes of the CO isotopologue lines are measured and compared with the grid of physical-chemical models of protoplanetary disks to estimate the disk gas mass \citep{Williams14, Miotello16}.
A  correlation between the disk gas mass and the stellar mass is also observed, 
but the correlation is less significant compared to that between the disk dust mass and the stellar mass due to the smaller number of the detections in the CO isotopologue lines and large uncertainties in deriving the total gas mass from line fluxes \citep{Ansdell16, Miotello17, Long17}. 
On the other hand, no correlation was found between the disk gas mass and the mass accretion rate, in contrast to the expectation from viscously evolving disks \citep{Manara16}. 
In addition, the gas-to-dust mass ratios derived with the disk gas mass estimated from the CO isotopologue lines are typically between 1 to 10 in these surveys \citep{Ansdell16, Miotello17, Long17}. 
These results hint at the uncertainty in estimating the disk gas mass with CO isotopologues, possibly due to the carbon depletion \citep[e.g.,][]{Miotello17}.  
Results of chemical models of protoplanetary disks also suggest that the disk gas mass can be underestimated by a factor of a few to two orders of magnitude with CO isotopologue lines when the conventional ISM CO abundance is adopted \citep{Miotello14, Miotello16, Yu16, Yu17, Molyarova17}.
Nevertheless, a larger sample of disks detected in molecular lines is needed to study the relations between disk gas and dust masses and mass accretion rate.
 
With an on-source integration time of a few minutes in the current ALMA surveys on protoplanetary disks, 
the gas distributions and velocity structures of the disks traced by the molecular lines can still not be imaged at high S/N ratios, except for a few very bright disks. 
In most of the disks, the CO isotopologue lines are detected after integrating the emission over spatial and velocity ranges  \citep{Ansdell16, Ansdell17, Long17}.
The central stellar mass and disk geometry of the majority of the disks in the surveys cannot be constrained from the gas kinematics because of the limited S/N ratios in the velocity channel maps. 
Thus, currently it is not possible to compare stellar masses dynamically determined from gas kinematics with spectroscopically determined masses in stellar evolutionary models of a large sample.
Such a comparison is, nevertheless, important to understand stellar evolution of young stellar objects \citep[e.g.,][]{Rizzuto16, Simon17}.
Applying further techniques, such as Keplerian masking, is required to better reveal distributions of molecular lines and to constrain gas kinematics \citep[e.g.,][]{Matra15, Marino16, Salinas17}. 

In order to extract more information from molecular-line data of these ALMA surveys and to study disk properties traced by molecular lines, 
we apply in this work the velocity-aligned stacking method described in \citet{Yen16} on the $^{13}$CO (3--2) and C$^{18}$O (3--2) data of the Lupus survey \citep{Ansdell16}. 
With this method, we enhance the S/N ratios of the data and obtain better detections of the disks in the $^{13}$CO and C$^{18}$O lines. 
We can then constrain disk orientation and stellar mass and re-estimate the disk gas mass with new measurements of the line flux.
This paper is organized as follows: Section 2 introduces the observations and the data. Section 3 describes and demonstrates our method to measure disk parameters in the $^{13}$CO (3--2) line. Section 4 presents our measurements of disk orientation, disk gas mass, and stellar mass. Section 5 discusses the relations between dynamical and spectroscopic stellar masses and between  the disk gas and dust masses and the mass accretion rate found in our results. 

\section{Observations}
The data analyzed here were retrieved from the ALMA archive (project code: 2013.1.00220.S).  
In this ALMA project, 88 young stellar objects (YSOs) having Class II or flat infrared spectra in the Lupus star-forming region were observed with an on-source integration time of approximately one minute per source.
0.9 mm continuum, $^{13}$CO (3--2; 330.587965 GHz), and C$^{18}$O (3--2; 329.330552 GHz) were observed simultaneously. 
The details of the observations have been described in \citet{Ansdell16}. 
In this work, we re-analyzed the $^{13}$CO and C$^{18}$O (3--2) data. 
The raw visibility data were calibrated using the standard reduction script for the cycle-2 data, which uses tasks in Common Astronomy Software Applications \citep[CASA;][]{McMullin07} of version 4.2.2. 
The image cubes of the $^{13}$CO and C$^{18}$O (3--2) lines were generated with the briggs weighting with a robust parameter of +0.5 and cleaned with the CASA task ``clean'' at a spectral resolution of 122 kHz, corresponding to a velocity resolution of 0.11 km s$^{-1}$. 
The typical angular resolution achieved is $\sim$0\farcs34, 
and the typical noise levels per channel are 70 and 67 mJy beam$^{-1}$ for the $^{13}$CO (3--2) and C$^{18}$O (3--2) emission, respectively.

\section{Method}\label{method}
We applied the velocity-aligned stacking method described in \citet{Yen16} on the ALMA $^{13}$CO and C$^{18}$O (3--2) data of the Lupus sample. 
This method aligns spectra at different positions in a protoplanetary disk by shifting them by the projected Keplerian velocities at their positions and then stacks them. 
With this alignment, the signals at the different positions are coherently added, and the total flux is accumulated in a narrower velocity range. 
As a result, the S/N ratio of an aligned stacked spectrum is enhanced. 
We adopted this method to enhance the S/N ratios of the ALMA $^{13}$CO and C$^{18}$O data and to estimate the stellar mass of the targets and the orientations of their associated disks. 

With a known stellar position and the assumption that its disk traced by the $^{13}$CO and C$^{18}$O (3--2) lines is geometrically thin, 
the velocity pattern of Keplerian rotation of this disk can be described with four parameters, stellar mass ($M_\star$), position angle of the major axis (PA), inclination angle ($i$), and systemic velocity ($V_{\rm sys}$). 
A detailed formulation is given in \citet{Yen16}.
In this work, for each source, we generated a series of aligned stacked spectra with different combinations of these four parameters. 
Each aligned stacked spectrum was generated by integrating emission over the entire disk area (i.e., azimuthally over 2$\pi$ and radially from the center to the disk outer radius).
Then, we searched for the parameter set that maximized the S/N ratios of the aligned stacked spectra, and adopted this parameter set as our measurements of $M_\star$, PA, $i$, and $V_{\rm sys}$ of that source.
In other words, we measured $M_\star$, PA, $i$, and $V_{\rm sys}$ of each source by maximizing the auto-correlation between the data and the various generated Keplerian rotational patterns. 

In our analysis, 
the stellar position of each source is adopted to be the center of its 0.9 mm continuum emission observed in the same observations or the pointing center of the observations if the continuum is not detected. 
The coordinates are obtained from \citet{Ansdell16}. 
The distance is adopted to be 200 pc for the sources in the Lupus III cloud and to be 150 pc for the sources in the Lupus I, II, and IV clouds \citep{Comeron08}.
The parameter ranges that we searched to maximize the S/N ratios of the aligned stacked spectra are $0.1\ M_\sun \leq M_\star \leq 3\ M_\sun$ in steps of $0.1\ M_\sun$ or an increase by 10\% when $M_\star \geq 1.5\ M_\sun$ , $0\degr \leq {\rm PA} \leq 355\degr$ in steps of 5$\degr$, $5\degr \leq i \leq 85\degr$ in steps of 5$\degr$, and $1.7\ {\rm km s}^{-1} \leq V_{\rm sys} \leq 6.3\ {\rm km s}^{-1}$ in steps of 0.1 km s$^{-1}$.
PA is defined as the direction from the blue- to redshifted emission.
From the stacked moment 0 map of the $^{13}$CO emission of all the sources detected in the $^{13}$CO (3--2) line made by \citet{Ansdell16}, 
we found that the $^{13}$CO emission is primarily within a radius of 1$\arcsec$.
Thus, in our parameter search, 
all the aligned stacked spectra were generated by integrating the area within a de-projected radius of 1$\arcsec$.  
%This radius is determined as the radius of the emission in the stacked moment 0 map of the $^{13}$CO emission of all the sources detected in the $^{13}$CO (3--2) line made by \citet{Ansdell16}. 
Our integrated area varies with $i$, as the projected area of a circular disk on the plane of the sky changes. 

As demonstrated in \citet{Yen16}, an aligned stacked spectrum is centered at $V_{\rm sys}$ and is approximately symmetric with respect to $V_{\rm sys}$, when data are properly aligned. 
Thus, for each aligned stacked spectrum, we fitted a Gaussian line profile. 
When the fitted line center is close to $V_{\rm sys}$ within the 1$\sigma$ Gaussian line width, 
we computed the S/N ratio of its integrated intensity within the 1$\sigma$ Gaussian line width. 
The integration was weighted by the fitted Gaussian line shape, $\int I_\nu \times G_\nu d\nu$/$\int G_\nu d\nu$, 
where $I_\nu$ is the observed spectrum and $G_\nu$ is the fitted Gaussian profile with the peak scaled to be one, such that channels closer to $V_{\rm sys}$ have larger weights. 
For a Gaussian-like spectrum, its weighted integrated intensity is proportional to its unweighted integrated intensity divided by its line width.
Thus, for aligned stacked spectra with the same total flux, the one that is more symmetric and has a narrower line width has a higher S/N ratio than the others. 
In addition, as shown in \citet{Yen16}, the noise of aligned stacked spectra changes with different alignments even if the integrated area is the same. 
This is caused by de-correlation of pixels within one synthesized beam due to the alignment. 
Hence, when we compared the S/N ratios of stacked spectra aligned with different parameters, their noise levels are computed from the original non-aligned data cube with the same area and channel ranges adopted to generate these aligned stacked spectra. 
Note that the area adopted to generate aligned stacked spectra changes with $i$, 
and the integrated channel range to compute S/N ratios depends on the line profiles of spectra. 
Better aligned spectra tend to have narrower line widths and higher integrated fluxes, leading to higher S/N ratios.
In this case, our comparison of the S/N ratios is not biased by the effects of the de-correlation. 

After the S/N ratios of the weighted integrated intensity of all the stacked spectra aligned with different parameters were computed, 
we calculated the S/N-ratio-weighted means and dispersions (i.e., moment 1 and 2) of PA, $i$, and $V_{\rm sys}$, 
and the number distribution of S/N ratios. 
Three examples of the distributions of the S/N ratios in the parameter space are presented in Appendix \ref{covar}.
The aligned stacked spectra with the S/N ratios of their weighted integrated intensity below 3$\sigma$ are excluded, 
and we trimmed the outliers in the number distribution at the high S/N-ratio end because they are found to be isolated in the parameter space and are false signals caused by shifting high-noise channels to $V_{\rm sys}$.  
Then, we searched for the maximum S/N ratio near the means and within the dispersions of the parameters. 
%With this process, false signals made by the alignment coincidently shifting more positive noise to $V_{\rm sys}$, 
%whose computed S/N ratios appeared to be more distinct from their neighbor parameters, 
%were rejected in the parameter search. 
The parameter set resulting in the highest S/N ratio is adopted as our measurements. 

To estimate the uncertainties of our measurements for each source, 
we searched for its stacked spectra that are aligned with different parameters but have the line profiles consistent with the best-aligned stacked spectrum within the uncertainty. 
For an aligned stacked spectrum, its line profile was quantified with its peak intensity, center, and width measured from Gaussian fitting. 
With these three quantities, the line profile of an aligned stacked spectrum generated with different parameters was compared with the best-aligned stacked spectrum. 
When their peak intensities, centers, and widths are all consistent within the uncertainties, 
and the difference in their S/N ratios of the integrated fluxes is less than $\sqrt{2}$, 
we claim that this spectrum is consistent with the best-aligned stacked spectrum. 
We adopted the ranges of the parameters resulting in the spectra consistent with the best-aligned stacked spectrum as the uncertainties of our measurements of that source. 

We performed this method on the $^{13}$CO (3--2) data of 88 sources observed with ALMA. 
A detection of the $^{13}$CO (3--2) emission is claimed if (1) the peak intensity of the best-aligned stacked spectrum is above 4$\sigma$ and (2) there are at least three channels above 3$\sigma$ within the 2$\sigma$ Gaussian line width. 
Then we applied the best parameters of the alignment on the C$^{18}$O (3--2) data because the S/N ratios of the C$^{18}$O (3--2) data are not sufficiently high to constrain the parameters even with this method.
We have tested the method on 30 synthetic ALMA images of a blank field, which were generated with the CASA simulator and have the same resolution and noise level as the observations. 
Our method indeed did not find detections in these synthetic images. 
Thus, our criteria of detection are sufficiently stringent to rule out the possibility that our method coherently adds noise and mimics a signal by coincidence. 

\subsection{Demonstration with observational data}\label{demoob}

\begin{figure*}
\centering
\includegraphics[width=18cm]{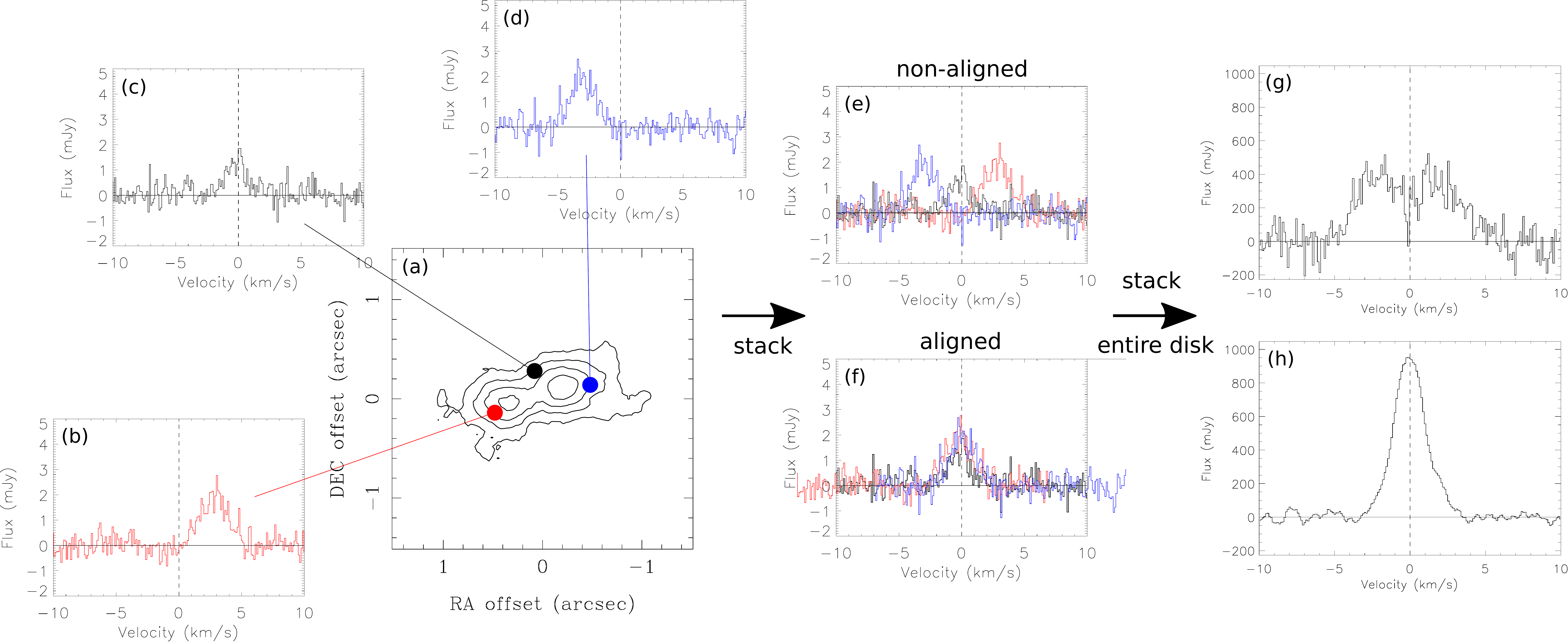}
\caption{Demonstration of our method with the $^{13}$CO (3--2) data of \object{J16083070$-$3828268}. (a) Moment 0 map of the $^{13}$CO (3--2) emission in \object{J16083070$-$3828268} obtained with the ALMA observations. (b)--(d) $^{13}$CO Spectra at the same de-projected radius of 0\farcs5 but at different position angles in the disk, shown as red, black, and blue dots in panel (a). (e) \& (f) Demonstration of stacking the three spectra with and without aligning them with the Keplerian velocities at their positions in the disk. (g) \& (h) Final stacked spectra integrated over the entire disk with and without the alignment.}\label{demo}
\end{figure*}

In this ALMA dataset, 
we found only few sources where the $^{13}$CO emission is sufficiently bright to directly image its distribution and velocity structures. 
In Fig.~\ref{demo}, we demonstrate our method with the  $^{13}$CO data of one of those bright sources, \object{J16083070$-$3828268}.
The $^{13}$CO spectra at different positions in the disk around \object{J16083070$-$3828268} are centered at different velocities (Fig.~\ref{demo}b--d) because of different projected Keplerian velocities at these positions.
Consequently, when these spectra are directly stacked together, the emission is not coherently added (Fig.~\ref{demo}e), resulting in a stacked spectrum with double peaks and a wide line width of $\sim$10 km s$^{-1}$ (Fig.~\ref{demo}g).
In contrast, our method aligns these spectra with different centroid velocities first before stacking them (Fig.~\ref{demo}f).
Thus, the emission originated from different positions in the disk is coherently added, and the stacked spectrum has a single peak and a narrower line width of $\sim$6 km s$^{-1}$.
The peak intensity also increases by almost a factor of two, and the S/N ratio of the stacked spectrum is enhanced.

\begin{table*}
\caption{Results of test with observational data}\label{test1}
\centering
\begin{tabular}{lcccccccccc}
\hline\hline
 & & \multicolumn{4}{c}{From P--V Diagram} && \multicolumn{4}{c}{From Aligned Spectrum} \\
 \cline{3-6} \cline{8-11}
Source & $M_{\star,{\rm spec}}$ & $M_\star$ & PA & $i$ & $V_{\rm sys}$ && $M_\star$ & PA & $i$ & $V_{\rm sys}$ \\
& ($M_\sun$) & ($M_\sun$) & (\degr) & (\degr) & (km s$^{-1}$) && ($M_\sun$) & (\degr) & (\degr) & (km s$^{-1}$) \\
\hline  \\
\object{RY~Lup} & 1.47$\pm$0.22 & 1.0$\pm$0.2 & 287$\pm$6 & 63$\pm$8 & 3.8$\pm$0.2 && 1.3$\pm$0.1 & 290$^{+ 5}_{-10}$ &  55$\pm$5 & 3.8$^{+0.3}_{-0.1}$ \smallskip \\ 
\object{J16083070$-$3828268} & 1.81$\pm$0.28 & 1.4$\pm$0.2 & 106$\pm$7 & 66$\pm$6 & 5.3$\pm$0.1 && 1.5$\pm$0.1 & 110$\pm$5 &  55$\pm$5 & 5.2$\pm$0.1 \smallskip \\
\object{Sz~83} & 0.75$\pm$0.19 & 0.5$\pm$0.6 & 116$\pm$80 & 19$\pm$12 & 4.6$\pm$0.1 && 0.2$^{+0.3}_{-0.1}$ & 120$^{+10}_{- 5}$ & 35$^{+ 5}_{-15}$ & 4.6$\pm$0.1  \smallskip \\
\hline
\end{tabular}
\tablefoot{$M_{\star,{\rm spec}}$ is the spectroscopically determined stellar mass from \citet{Alcala14, Alcala17}. PA is defined as the direction from the blue- to redshifted emission.}
%\tablebib{(1) \citet{branch83}; (2) \citet{phillips87}; (3) \citet{barbon90}}
\end{table*}

\begin{figure*}
\centering
\includegraphics[width=17cm]{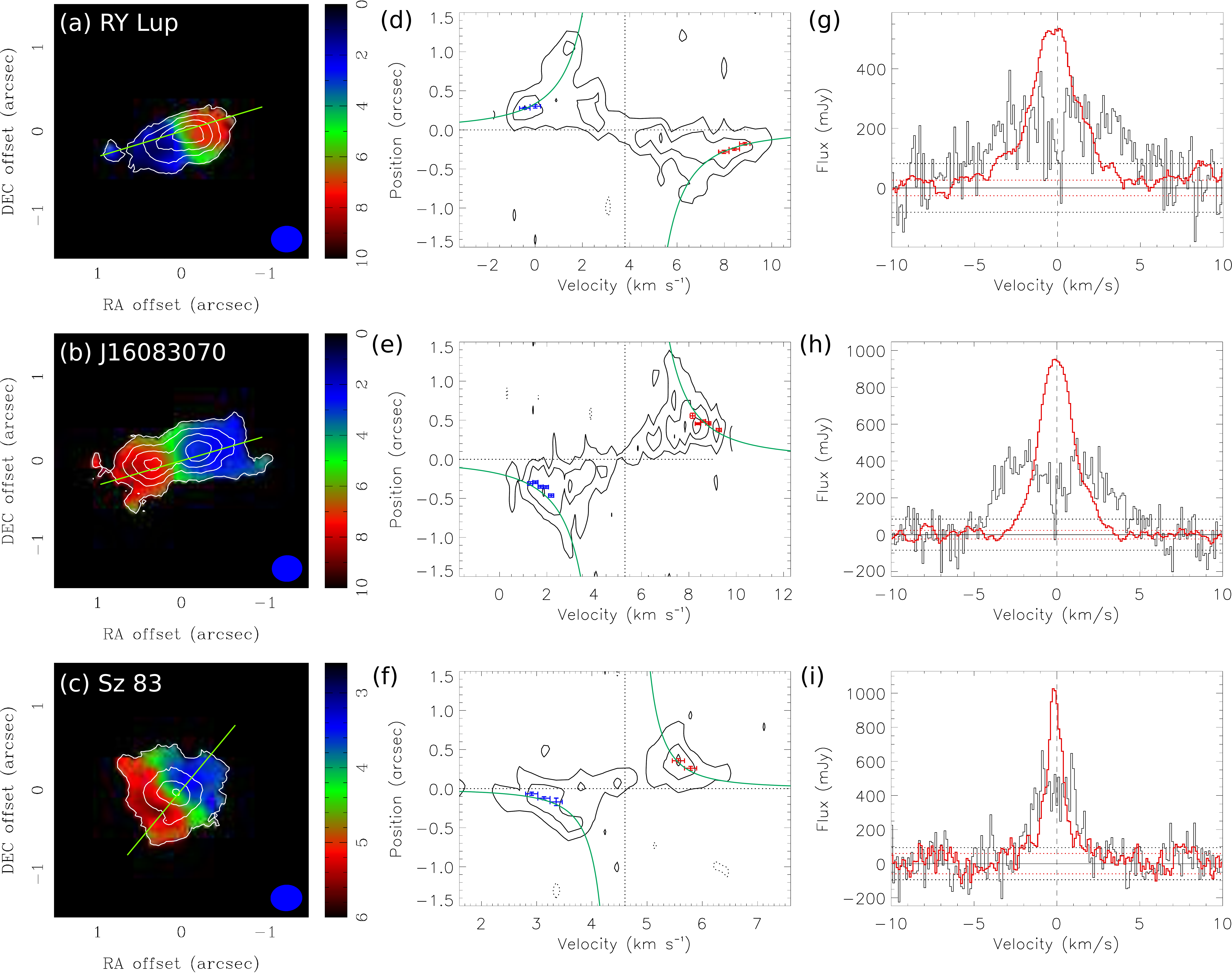}
\caption{ALMA observational results of \object{RY~Lup} (top row), \object{J16083070$-$3828268} (middle row), and \object{Sz~83} (bottom row). (a)--(c) Moment 0 map (contour)  overlaid on the moment 1 map (color; in units of km s$^{-1}$) of the $^{13}$CO emission. Green lines denote the axes where the PV diagrams are extracted. Contour levels start from 3$\sigma$ in steps of 3$\sigma$, where 1$\sigma$ is 77 mJy km s$^{-1}$ in (a), 74 mJy km s$^{-1}$ in (b), and 44 mJy km s$^{-1}$ in (c). (d)--(f) PV diagrams of the $^{13}$CO emission along the major axes and passing through the stellar positions. Blue and red data points denote the measured peak positions in the high-velocity channels, and green curves present the fitted Keplerian rotational profiles to the data points. Contour levels start from 2$\sigma$ in steps of 2$\sigma$, where 1$\sigma$ is 35 mJy km s$^{-1}$ in (d) and 50 mJy in (e) and (f). (g)--(i) Stacked $^{13}$CO spectra with (red) and without (black) alignment integrated over the disk area. Zero velocity refers to the measured $V_{\rm sys}$. Red and black horizontal dotted lines denote $\pm$1$\sigma$ levels with and without alignment, respectively.}\label{data_map}
\end{figure*}

For the bright sources in this Lupus sample, we also analyzed their data with conventional methods to test the robustness of our velocity-aligned stacking method. 
Below, we present three cases, \object{RY~Lup}, \object{J16083070$-$3828268}, and \object{Sz~83}.
Figure \ref{data_map}a--c presents the total integrated intensity (moment 0) and intensity-weighted mean velocity (moment 1) maps of the $^{13}$CO emission in \object{RY~Lup}, \object{J16083070$-$3828268}, and \object{Sz~83}. 
The $^{13}$CO components in \object{RY~Lup} and \object{J16083070$-$3828268} are elongated along the northwest--southeast direction, 
and exhibit clear velocity gradients along the elongation. 
The $^{13}$CO component in \object{Sz~83} also exhibits a clear velocity gradient but does not show any obvious elongation.
We fitted a two-dimensional Gaussian function to the intensity distributions of the $^{13}$CO emission. 
The deconvolved full-width-half-maximum (FWHM) size is measured to be 0\farcs85$\pm$0.11 $\times$ 0\farcs34$\pm$0.04 with a PA of the major axis of 287\degr$\pm$6\degr in \object{RY~Lup}. 
Those in \object{Sz~83} are measured to be 0\farcs7$\pm$0.11 $\times$ 0\farcs66$\pm$0.04 and 116\degr$\pm$80\degr, respectively. 
The PA of the major axis in \object{Sz~83} is poorly constrained because the aspect ratio is almost unity. 
The $^{13}$CO emission in \object{J16083070$-$3828268} shows two peaks, so its intensity distribution cannot be fitted with a two-dimensional Gaussian function. 
The PA of the major axis in \object{J16083070$-$3828268} is measured from the axis passing through the two peaks to be 106\degr$\pm$7\degr. 
The size of the intensity distribution is 1\farcs27$\pm$0\farcs03 $\times$ 0\farcs52$\pm$0\farcs02 from the FWHM widths of the intensity profiles extracted along and perpendicular to the elongation. 

We extracted the position--velocity (PV) diagrams along the major axes and passing through the continuum peak positions in these sources (Fig.~\ref{data_map}d--f). 
The PV digram of \object{Sz~83} was extracted along the PA of the velocity gradient, 140\degr. 
The data cube of \object{RY~Lup} was binned up every four channels, and those of  \object{J16083070$-$3828268} and \object{Sz~83} were binned up every two channels to increase the S/N ratios per channel. 
We measured the peak positions at different velocity channels in the PV diagrams following the method described in \citet{Yen13}.
The distances from the peak positions to the stellar positions were adopted as rotational radii ($R_{\rm rot}$), 
and the relative velocities with respect to $V_{\rm sys}$ at those velocity channels were adopted as rotational velocities ($V_{\rm rot}$). 
We fitted Keplerian rotational profiles, $V_{\rm rot} \propto {R_{\rm rot}}^{-0.5}$, to these data points obtained from the PV diagrams, 
and there were two free parameters in the fitting, $V_{\rm rot}$ at a representative radius and $V_{\rm sys}$. 
The measured Keplerian velocities are 2.2$\pm$0.1 km s$^{-1}$ at a radius of 150 AU in \object{RY~Lup}, 2.3$\pm$0.1 km s$^{-1}$ at a radius of 200 AU in \object{J16083070$-$3828268}, and 0.56$\pm$0.04 km s$^{-1}$ at a radius of 150 AU in \object{Sz~83}.
On the assumption that the disks around these sources observed in the $^{13}$CO emission are circular and geometrically thin, 
the inclination angles, $i$, are estimated to be 63\degr$\pm$8$\degr$ in \object{RY~Lup}, 66\degr$\pm$6$\degr$ in \object{J16083070$-$3828268}, and 19\degr$\pm$12$\degr$ in \object{Sz~83} from the aspect ratios of the major and minor axes.
Then, with the measured Keplerian velocities and inclination angles, we derived $M_\star$ of \object{RY~Lup}, \object{J16083070$-$3828268}, and \object{Sz~83}. 
The estimated $M_\star$, PA, $i$, $V_{\rm sys}$ in these three sources with the analysis described above are listed in Table \ref{test1}.
We note that \object{Sz~83} is close to face on, so its estimated $M_\star$, which is proportional to $1/\sin i^2$, is very uncertain.

We also measured the $M_\star$, PA, $i$, $V_{\rm sys}$ of these three sources with our velocity-aligned stacking method. 
The aligned stacked spectra with maximum S/N ratios are shown in Fig.~\ref{data_map}g--i.
The measurements from the velocity-aligned stacking method are also listed in Table \ref{test1} for comparison. 
The distributions of the highest S/N ratio achieved with a given pair of parameters are presented in Appendix \ref{covar} to demonstrate the covariance between these parameters.
We found that there is a clear correlation in the S/N ratio distribution between $M_\star$ and $i$. 
Thus, the uncertainty in $i$ can propagate to $M_\star$, resulting in poorly constrained $M_\star$, as in the case of \object{Sz~83}.  
On the other hand, no correlation is found between other pairs of parameters, so there is no significant covariance between these parameters.
This demonstration of our velocity-aligned stacking method with the observational data shows that 
the results from the conventional method and our new method are consistent within the uncertainties, 
and our new method indeed provides robust measurements of $M_\star$, PA, $i$, and $V_{\rm sys}$ with high S/N-ratio data.

\subsection{Demonstration with models}\label{demo2}

\begin{figure*}
\centering
\includegraphics[width=17cm]{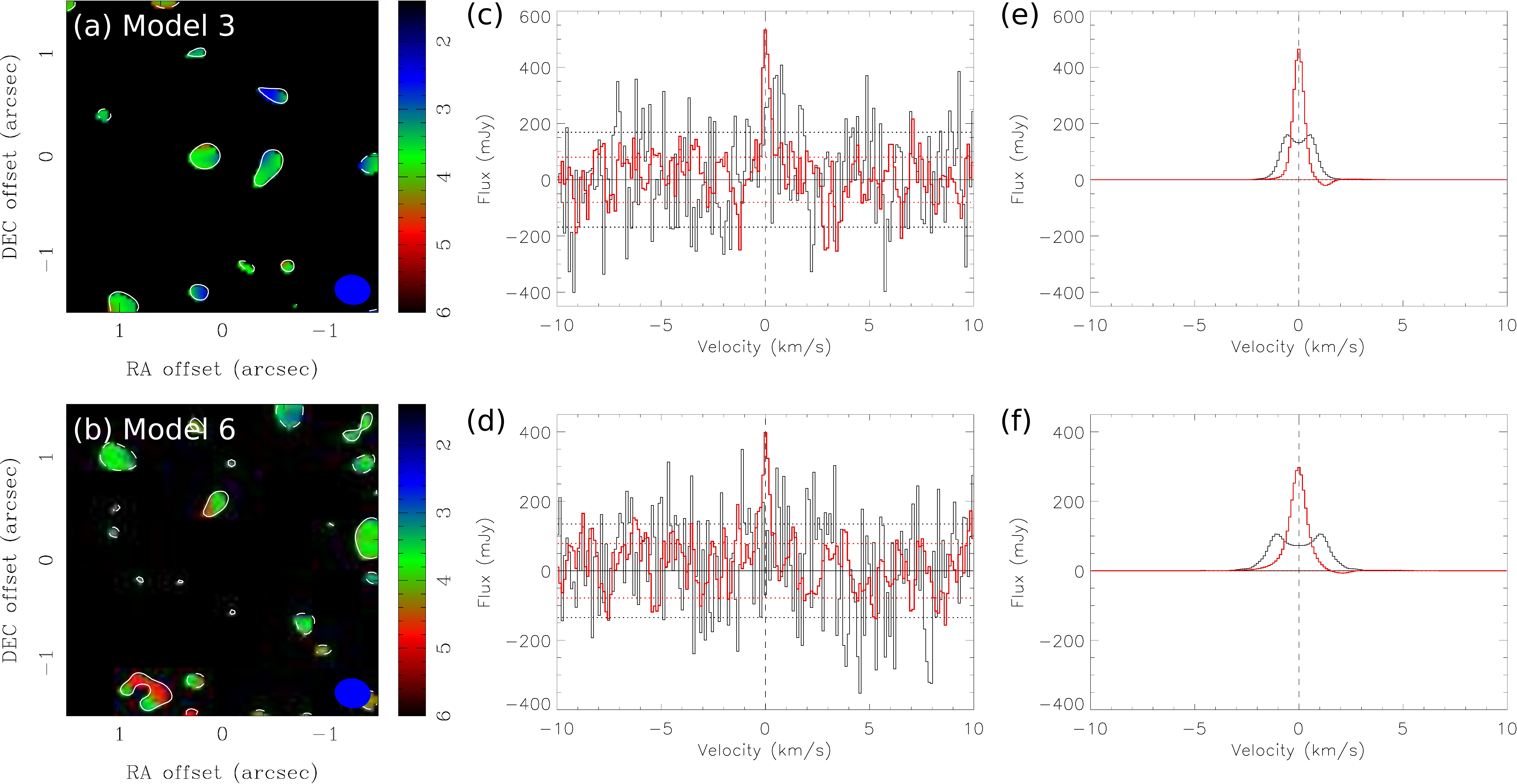}
\caption{Synthetic images and spectra of disk model 3 (top row) and 6 (bottom row). The model parameters are listed in Table \ref{test2}. (a) \& (b) Moment 0 maps (contour)  overlaid on the moment 1 maps (color; in units of km s$^{-1}$) of the model disks. The moment 0 and 1 maps were computed by integrating the velocity range of $V_{\rm LSR} = 3.7\pm2.3$ km s$^{-1}$, the same as in \citet{Ansdell16}. Contour levels start from 2$\sigma$ in steps of 2$\sigma$, where 1$\sigma$ is 56 mJy km s$^{-1}$. (c) \& (d) Synthetic stacked spectra with the alignment using the measured parameters (red) overlaid on the ones without alignment (black). Both are integrated over the disk area. Zero velocity refers to the measured $V_{\rm sys}$. Red and black horizontal dotted lines denote $\pm$1$\sigma$ levels with and without alignment, respectively. (e) \& (f) Same as (c) \& (d) but for synthetic spectra without noise. Black histograms present the stacked spectrum without alignment, and red histograms present the stacked spectrum with the alignment using the input parameters.}\label{model_map}
\end{figure*}

To test our method on faint disks, we generated a series of synthetic images of simple model disks. 
The model disks were made with different intensity distributions, stellar masses, inclination angles, and total integrated fluxes. 
The details of the model generation are described in Appendix \ref{mkmod}, and the model parameters are listed in Table \ref{test2}.
Then, we simulated ALMA observations with a one-minute integration time on our model image cubes.  
Our synthetic images have the same angular resolution and noise level as the observations. 
We applied our velocity-aligned stacking method on the synthetic data to measure $M_\star$, PA, $i$, and $V_{\rm sys}$ of these model disks.
The measurements from our method are listed and compared with model inputs in Table \ref{test2}. 
In this test, we detected 17 out of 24 model disks. 
The detailed results and discussion of this demonstration with the model disks are presented in Appendix \ref{detail}.
Figure~\ref{model_map} presents two examples of the synthetic moment 0 and 1 maps and stacked spectra with and without alignment of our model disks. 
There is no clear detection in neither the moment 0 maps nor the stacked spectra without alignment. 
For comparison, we also generated synthetic spectra without simulating observational noise (Fig.~\ref{model_map}e \& f). 
As shown in the synthetic spectra without noise, the intensity of the model disks is below the 1$\sigma$ noise level of the observations, %synthetic spectra, 
and the intensity becomes higher than the noise level after applying the alignment (Fig.~\ref{model_map}e \& f).
Thus, after applying our method, the detections are obtained in the synthetic aligned stacked spectra. 
This test demonstrates that even if the intensity distribution and velocity structure of a disk are not detected in its moment 0 and 1 maps and spectrum, our method can still find the signals and measure its $M_\star$, PA, $i$, and $V_{\rm sys}$. 
The results of our test show that we are able to detect disks with a total integrated flux as low as 400 mJy km s$^{-1}$.  
The detection limit changes with $M_\star$, $i$, and intensity profiles, and it can be even lower than 400 mJy km s$^{-1}$. 
 
The measurements from the synthetic images are typically consistent with the model inputs within the 1$\sigma$--2$\sigma$ uncertainties, with exceptions where the difference is more than 3$\sigma$, e.g., model 22 and 23 (Table \ref{test2}).
The inclination is more difficult to constrain because it requires more spatial information along the direction of the minor axis. 
The detailed comparison between the measurements from the synthetic images and the model inputs is described in Appendix \ref{detail}. 
This demonstration with the synthetic data shows that our method can provide robust measurements for disks with an integrated line flux of $\gtrsim$400 mJy km s$^{-1}$. 
%The results of this test show that our method can provide robust measurements even for faint disks (Table \ref{test2}). 
However, when the total flux is as low as 400 mJy km s$^{-1}$, $M_\star$ can be overestimated because of the limitation in constraining a disk orientation (e.g., model 22 in Table \ref{test2}).
As discussed in Section~\ref{mstar}, these possibly overestimated $M_\star$ can be identified from their flux densities. 
Nevertheless, that does not affect the robustness of the measured flux because our method conserves the flux \citep{Yen16}. 
In this demonstration, on average our method detects 80\% of the total fluxes of the model disks (Table \ref{test2}). 
There are only two cases where only 50\% of the total fluxes are detected, and 
only one case where the measured flux is 70\% higher than the input flux. 
Thus, the uncertainty in the measured flux due to the limited sensitivity with this method is typically 20\% and exceptionally it can be a factor of two. 

\section{Measured Stellar Mass and Disk Properties}

\begin{table*}
\caption{Stellar mass and properties of detected disks from the velocity-aligned stacking method.}\label{measure}
\centering
\begin{tabular}{lccccccc}
\hline\hline
Source & $M_\star$ & PA & $i$ & $V_{\rm sys}$ & $F_{\rm ^{13}CO}$ & $F_{\rm C^{18}O}$ & $R_{\rm det}$ \\
 & ($M_\sun$) & (\degr) & (\degr) & (km s$^{-1}$ &  (mJy km s$^{-1}$) & (mJy km s$^{-1}$) & (AU) \\
\hline \\
Sz~65 & $>$0.8 & 295$^{+20}_{-10}$ & $<$50 & 4.5$^{+0.4}_{-0.6}$ &  992$\pm$53 & $<$291 & 170 \smallskip\\
J15450887$-$3417333 & 0.1$\pm$0.1 & 170$^{+10}_{-40}$ & 60$^{+ 5}_{-15}$ & 4.5$^{+0.4}_{-0.2}$ &  618$\pm$40 &  107$\pm$11 & 150 \smallskip\\
Sz~68\tablefootmark{a} & $>$1.6 & 110$^{+10}_{- 5}$ & $<$45 & 2.6$^{+0.2}_{-0.1}$ &  434$\pm$28 &  246$\pm$28 & 120 \smallskip\\
Sz~69 & $>$0.4 & 315$\pm$10 & $<$40 & 5.4$^{+0.5}_{-0.2}$ &  598$\pm$53 & $<$216 & 140 \smallskip\\
Sz~71 & $>$0.7 &  35$^{+10}_{- 5}$ & $<$30 & 3.6$\pm$0.1 & 1394$\pm$117 & $<$433 & 240 \smallskip\\
Sz~72\tablefootmark{a}  & 0.5$\pm$0.1 & 315$\pm$5 &  75$\pm$5 & 2.9$^{+0.3}_{-0.2}$ &  119$\pm$10 & $<$ 44 & 110 \smallskip\\
Sz~73\tablefootmark{a}  & $>$1.3 & 255$^{+ 5}_{-10}$ & $<$50 & 3.6$^{+0.3}_{-0.2}$ &  351$\pm$22 &  136$\pm$24 & 120 \smallskip\\
Sz~83 & 0.2$^{+0.3}_{-0.1}$ & 120$^{+10}_{- 5}$ & 35$^{+ 5}_{-15}$ & 4.6$\pm$0.1 & 2600$\pm$96 &  671$\pm$57 & 170 \smallskip\\
Sz~84 & 0.9$^{+0.6}_{-0.2}$ & 355$^{+ 5}_{-**}$ & 50$^{+ 5}_{-15}$ & 4.8$^{+0.4}_{-0.3}$ &  682$\pm$36 &  165$\pm$19 & 140 \smallskip\\
Sz~129 & 0.4$^{+0.1}_{-0.2}$ & 170$^{+40}_{-10}$ & 70$^{+ 5}_{-10}$ & 3.5$^{+0.3}_{-0.4}$ &  250$\pm$34 & $<$124 & 190 \smallskip\\
RY~Lup & 1.3$\pm$0.1 & 290$^{+ 5}_{-10}$ &  55$\pm$5 & 3.8$^{+0.3}_{-0.1}$ & 4368$\pm$84 & 1162$\pm$44 & 220 \smallskip\\
J16000236$-$4222145 & 1.0$^{+1.2}_{-0.2}$ & 340$\pm$5 & 30$^{+ 5}_{-10}$ & 4.1$\pm$0.1 & 1918$\pm$86 &  430$\pm$96 & 190 \smallskip\\
Sz~130 & 0.3$^{+0.2}_{-0.1}$ & 325$^{+10}_{-20}$ & 55$^{+10}_{-15}$ & 4.4$^{+0.3}_{-0.2}$ &  284$\pm$28 &  163$\pm$26 & 140 \smallskip\\
MY~Lup & 0.9$^{+0.2}_{-0.1}$ & 200$^{+10}_{- 5}$ &  55$\pm$5 & 4.1$^{+0.2}_{-0.1}$ & 1006$\pm$68 &  580$\pm$44 & 170 \smallskip\\
J16011549$-$4152351 & $>$0.5 & 310$^{+10}_{- 5}$ & $<$25 & 3.9$\pm$0.1 & 4418$\pm$169 & 1892$\pm$197 & 360 \smallskip\\
Sz~133 & 1.1$^{+1.1}_{-0.1}$ & 320$^{+10}_{- 5}$ & 50$^{+ 5}_{-20}$ & 5.1$^{+0.4}_{-0.1}$ &  487$\pm$31 &   82$\pm$8 & 150 \smallskip\\
Sz~88A\tablefootmark{a}  & 2.0$^{+0.4}_{-0.2}$ &  60$^{+ 5}_{-10}$ & 50$^{+ 5}_{-10}$ & 1.9$\pm$0.2 &  173$\pm$23 & $<$125 & 220 \smallskip\\
J16070384$-$3911113 & 0.3$^{+0.2}_{-0.1}$ & 340$\pm$10 & 50$^{+ 5}_{-15}$ & 3.1$^{+0.1}_{-0.2}$ & 1369$\pm$61 &  332$\pm$74 & 250 \smallskip\\
J16070854$-$3914075 & 0.6$^{+0.7}_{-0.1}$ & 345$^{+10}_{-20}$ & 50$^{+ 5}_{-20}$ & 2.7$^{+0.3}_{-0.2}$ & 1146$\pm$74 & $<$463 & 300 \smallskip\\
Sz~90 & 0.3$\pm$0.1 & 130$\pm$5 &  60$\pm$5 & 5.4$\pm$0.1 &  225$\pm$27 & $<$102 & 200 \smallskip\\
Sz~95 & 0.3$^{+0.3}_{-0.1}$ &  75$\pm$10 & 50$^{+ 5}_{-15}$ & 3.1$^{+0.1}_{-0.2}$ &  255$\pm$31 & $<$140 & 190 \smallskip\\
Sz~96\tablefootmark{a}  & $>$2.9 &  25$\pm$5 & $<$50 & 4.7$\pm$0.1 &  202$\pm$22 & $<$151 & 260 \smallskip\\
J16081497$-$3857145 & 0.4$\pm$0.1 &  35$^{+10}_{- 5}$ &  75$\pm$5 & 4.2$\pm$0.2 &  225$\pm$19 & $<$ 81 & 220 \smallskip\\
Sz~98 & 0.8$^{+0.8}_{-0.1}$ & 110$^{+10}_{-20}$ & 50$^{+ 5}_{-15}$ & 3.2$^{+0.3}_{-0.4}$ &  906$\pm$85 & $<$445 & 370 \smallskip\\
Sz~100 & 0.4$\pm$0.1 & 250$^{+ 5}_{-20}$ & 55$^{+ 5}_{-20}$ & 1.8$^{+0.2}_{-0.1}$ & 1033$\pm$41 &   66$\pm$15 & 220 \smallskip\\
Sz~103\tablefootmark{a}  & 0.8$^{+0.3}_{-0.1}$ &  95$\pm$5 & 50$^{+ 5}_{-10}$ & 2.1$\pm$0.1 &  138$\pm$17 & $<$101 & 190 \smallskip\\
J16083070$-$3828268 & 1.5$\pm$0.1 & 110$\pm$5 &  55$\pm$5 & 5.2$\pm$0.1 & 5983$\pm$87 & 1536$\pm$68 & 330 \smallskip\\
V856~Sco & 0.7$^{+1.1}_{-0.1}$ & 330$^{+10}_{-20}$ & 40$^{+ 5}_{-15}$ & 4.8$^{+0.1}_{-0.4}$ &  457$\pm$48 & $<$190 & 230 \smallskip\\
Sz~108B & 0.4$^{+0.4}_{-0.1}$ & 160$^{+20}_{- 5}$ & 50$^{+ 5}_{-20}$ & 2.5$\pm$0.1 &  280$\pm$27 &  136$\pm$22 & 160 \smallskip\\
J16085373$-$3914367\tablefootmark{a}  & 1.5$^{+0.1}_{-0.5}$ & 305$^{+20}_{- 5}$ &  65$\pm$5 & 2.2$^{+0.4}_{-0.2}$ &  234$\pm$19 & $<$ 72 & 190 \smallskip\\
Sz~111 & $>$1.2 &  40$^{+10}_{- 5}$ & $<$35 & 4.1$\pm$0.1 & 3448$\pm$173 &  517$\pm$42 & 430 \smallskip\\
J16090141$-$3925119 & 0.5$\pm$0.1 &  -5$\pm$5 &  60$\pm$5 & 3.4$^{+0.2}_{-0.1}$ & 1949$\pm$51 & $<$265 & 250 \smallskip\\
Sz~114 & 0.8$^{+1.0}_{-0.3}$ & 170$^{+ 5}_{-10}$ &  15$\pm$5 & 5.0$^{+0.1}_{-0.2}$ &  861$\pm$75 &  335$\pm$66 & 250 \smallskip\\
J16092697$-$3836269 & 0.2$\pm$0.1 & 130$^{+20}_{-10}$ &  65$\pm$5 & 4.1$\pm$0.2 &  239$\pm$26 & $<$117 & 200 \smallskip\\
J160934.2$-$391513\tablefootmark{a}  & 1.5$^{+0.5}_{-0.1}$ & 145$\pm$10 & 55$^{+ 5}_{-10}$ & 2.4$^{+0.1}_{-0.2}$ &  253$\pm$25 & $<$168 & 230 \smallskip\\
J16093928$-$3904316\tablefootmark{a}  & 0.9$\pm$0.1 &  65$\pm$5 &  75$\pm$5 & 3.3$\pm$0.1 &  145$\pm$17 & $<$ 82 & 270 \smallskip\\
Sz~118 & 1.0$^{+0.3}_{-0.2}$ & 155$^{+10}_{- 5}$ & 55$^{+ 5}_{-15}$ & 3.1$^{+0.4}_{-0.3}$ &  971$\pm$54 & $<$209 & 230 \smallskip\\
J16100133$-$3906449\tablefootmark{a}  & $>$2.2 & 190$^{+10}_{- 5}$ & $<$35 & 4.4$^{+0.1}_{-0.2}$ &  368$\pm$46 & $<$159 & 230 \smallskip\\
J16101984$-$3836065\tablefootmark{a}  & 1.6$^{+0.8}_{-0.1}$ & 335$^{+10}_{- 5}$ & 55$^{+ 5}_{-10}$ & 3.3$^{+0.1}_{-0.3}$ &  183$\pm$21 & $<$108 & 220 \smallskip\\
J16102955$-$3922144 & 0.2$\pm$0.1 & 120$^{+10}_{- 5}$ & 65$^{+ 5}_{-10}$ & 3.5$\pm$0.2 &  793$\pm$53 & $<$215 & 220 \smallskip\\
Sz~123A & 0.6$^{+0.9}_{-0.1}$ & 165$^{+10}_{-20}$ & 40$^{+ 5}_{-15}$ & 4.1$^{+0.3}_{-0.1}$ & 1212$\pm$61 &  307$\pm$28 & 190 \smallskip\\
\hline
\end{tabular}
\tablefoot{$M_\star$ is the stellar mass. PA, $i$, and $V_{\rm sys}$ are the position angle of the major axis, inclination angle, systemic velocity of the disk, respectively. $F_{\rm ^{13}CO}$ and $F_{\rm C^{18}O}$ are the integrated fluxes of the $^{13}$CO (3--2) and C$^{18}$O (3--2) emission, and $R_{\rm det}$ is the detected radius in the $^{13}$CO emission. The uncertainty in the integrated flux listed here only includes the noise of the data. There can be a systematic uncertainty of 20\% due to the limited sensitivity, as discussed in Section \ref{method}. \\
 \tablefoottext{a}{Source with a low $^{13}$CO flux density, where $M_\star$ can be overestimated (see Section \ref{mstar}).}}
%\tablebib{(1) \citet{branch83}; (2) \citet{phillips87}; (3) \citet{barbon90}}
\end{table*}

\begin{figure}
\centering
\includegraphics[width=7cm]{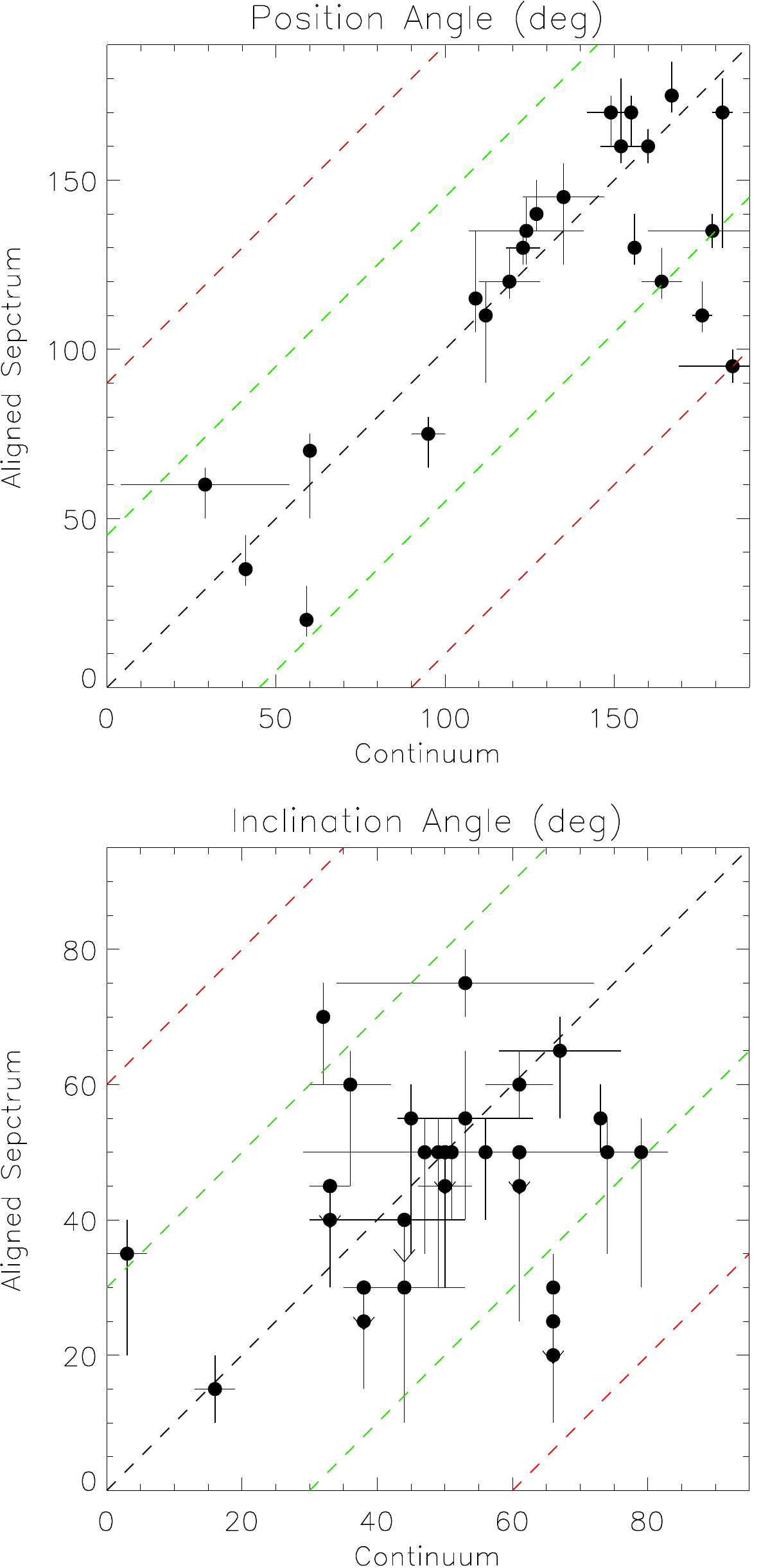}
\caption{Comparison of position angles (upper) and inclination angles (lower) estimated from the continuum emission (horizontal axis) in the literature and from the $^{13}$CO emission (vertical axis) with our method. In the upper panel, green and red dashed lines denote an angle difference of $\pm$45$\degr$ and $\pm$90$\degr$, respectively. In the lower panel, green and red dashed lines denote an angle difference of $\pm$30$\degr$ and $\pm$60$\degr$, respectively. }\label{corrdisk}
\end{figure}

With the velocity-aligned stacking method described in Section \ref{method}, we detect 41 out of 88 disks in the $^{13}$CO emission. 
11 of them were not detected in the $^{13}$CO emission in \citet{Ansdell16}. 
There are six disks detected in the $^{13}$CO emission in \citet{Ansdell16} which could not be identified in our analysis. 
Our method could not find parameters to align their data and obtain a detection at more than a 4$\sigma$ level in aligned stacked spectra.   
We note that our criteria to claim a detection are more stringent than in \citet{Ansdell16}, 
where they are requesting the integrated intensity to be larger than 3$\sigma$ level. 
All the measured $M_\star$, PA, $i$, and $V_{\rm sys}$ of the 41 disks are listed in Table \ref{measure}, and their velocity-aligned stacked spectra are shown in Appendix \ref{allspec}.
As demonstrated in Section \ref{demo2}, $M_\star$ can be overestimated when the integrated flux is lower than 400 mJy km s$^{-1}$, 
and those sources with possibly overestimated $M_\star$ are identified in Section \ref{mstar} and are labeled in Table \ref{measure}. 

Among our detected disks in $^{13}$CO, there are 23 disks which are resolved with their PA and $i$ also measured from the continuum emission \citep{Ansdell16, Tazzari17}.
Figure \ref{corrdisk}a and b present the comparison of PA and $i$ measured from our method and from the continuum emission. 
There is a clear correlation in PA from the two different methods, except for two outliers where the difference in PA is more than 45\degr. 
$i$ from the two methods are also correlated but have a larger scatter, as the uncertainties in $i$ are larger.  
The mean difference of the disk inclination, $i$, measured from our method and from the continuum emission is 15\degr, and 
$i$ for 18 out of 23 disks are consistent within 25\degr.
This comparison shows that our method detects independently and consistently the disk orientations, 
and hence, can also reveal a disk orientation in the absence of a clear direct detection in continuum or lines. 

\begin{figure}
\centering
\includegraphics[width=7cm]{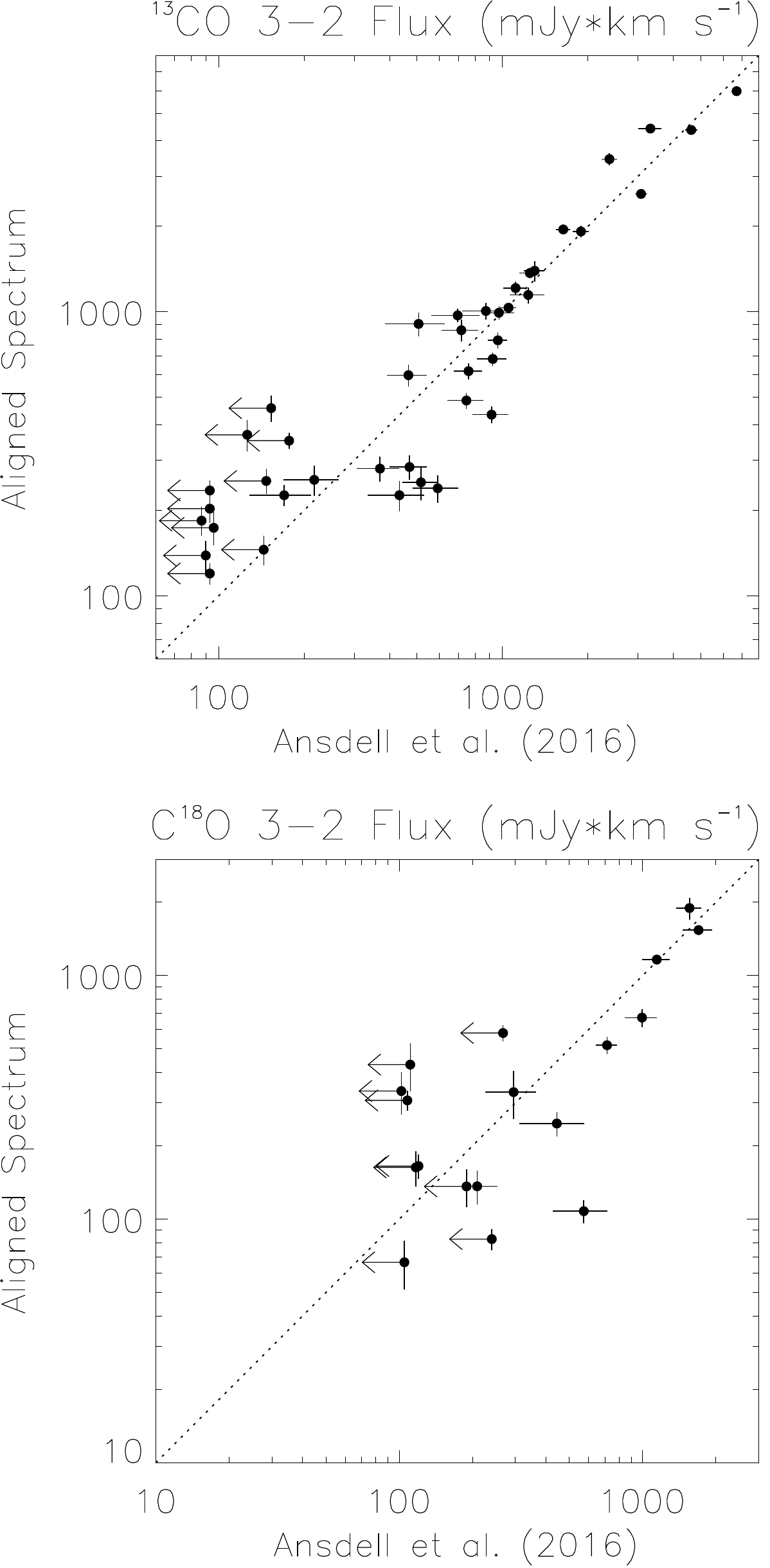}
\caption{Comparison of integrated $^{13}$CO (3--2) and C$^{18}$O (3--2) fluxes from our method (vertical axis) and in the literature (horizontal axis; \citet{Ansdell16}).}\label{corrflux}
\end{figure}

With the measured $M_\star$, PA, $i$, and $V_{\rm sys}$, 
for each source, we then generated aligned stacked spectra of the $^{13}$CO emission in a series of radial bins in steps of half of the synthesized beam.
We fitted Gaussian line profiles to these spectra and identified the outer radius where the detection was below 2$\sigma$. 
We adopted that radius as the detected disk radius of that source.  
Then, we generated the aligned stacked spectrum of the $^{13}$CO emission integrated over the area within the detected disk radius, 
and we measured the total integrated flux within twice of the FWHM line widths in the aligned stacked spectra.  
We applied the same parameters to align the C$^{18}$O data and generated aligned stacked spectra. 
Except for a few bright disks, the C$^{18}$O emission is still not clearly identified in the aligned stacked spectra.  
Nevertheless, we assume that the $^{13}$CO and C$^{18}$O lines trace the same regions, 
and we adopted the same area and velocity range as the $^{13}$CO emission to measure the total integrated flux of the C$^{18}$O emission.   
Although the distribution of C$^{18}$O is affected by isotope-selective effects more than $^{13}$CO \citep{Miotello16}, 
our selected spatial and velocity ranges for the integration from the $^{13}$CO emission are expected to fully cover the distribution of the C$^{18}$O emission.
We define that the C$^{18}$O emission is detected when the integrated flux is above the 4$\sigma$ threshold.
There are 18 disks detected in the C$^{18}$O line, and 9 of them are new detections compared to \cite{Ansdell16}.
Their velocity-aligned stacked spectra are shown in Fig.~\ref{spec2} and \ref{spec5}.
The detected disk radii and the measured integrated $^{13}$CO and C$^{18}$O fluxes are listed in Table \ref{measure}, 
and the 4$\sigma$ upper limit of the integrated C$^{18}$O flux is listed for the C$^{18}$O non-detected disks. 

%Figure \ref{corrdisk}c compares the outer disk radii measured from the continuum emission \citep{Ansdell16, Tazzari17} and our detected disk radii in the $^{13}$CO emission. 
%The outer disk radii in the continuum emission are defined as the radii containing 95\% of the total flux \citep{Tazzari17} or as twice the 1$\sigma$ width measured from %the two-dimensional Gaussian fitting to the visibility data \citep{Ansdell16}, which is also expected to contain 95\% of the total flux.
%The disk radii in the $^{13}$CO emission are all larger than those in the continuum as expectation from higher optical depth of the $^{13}$CO line \citep{Facchini17}.
Figure \ref{corrflux} compares our measured integrated fluxes of the $^{13}$CO and C$^{18}$O emission with the fluxed reported in \citet{Ansdell16}. 
Our measured fluxes are tightly correlated with the measurements in \citet{Ansdell16}.
Especially, for bright disks with integrated fluxes larger than 1000 mJy km s$^{-1}$, the difference between the fluxes from our method and from \citet{Ansdell16} ranges from a few percent to 30\%, and the mean difference is 10\%. 
On the other hand, for faint disks, the difference in the measured fluxes ranges from 10\% to as large as a factor of five, 
and several more disks with integrated fluxes of 100--500 mJy km s$^{-1}$ are detected with our method. 
We note that the area and velocity ranges for integration to measure the integrated fluxes are different between our method and \citet{Ansdell16}. 
In \citet{Ansdell16}, for disks that are not detected in the velocity channel maps, the integrated velocity ranges were all adopted to be $V_{\rm LSR}$ of 1.4--6 km s$^{-1}$, and the integrated area was determined by a curve-of-growth method on moment 0 maps. 
In contrast, our measurements show that $V_{\rm LSR}$ of the detected disks range from $V_{\rm LSR}$ of 1.8 to 5.4 km s$^{-1}$ (Table \ref{measure}). 
With a typical line width of a few km s$^{-1}$ of Keplerian disks (e.g., Fig.~\ref{data_map}), 
part of the emission can be located beyond $V_{\rm LSR}$ of 1.4--6 km s$^{-1}$ and is, hence not included in the calculation by \citet{Ansdell16}.
In addition, our method is expected to exclude more velocity channels with only noise in the integration because the integrated pixels and channels are selected based on a derived Keplerian rotation. 

\begin{figure}
\centering
\includegraphics[width=8.5cm]{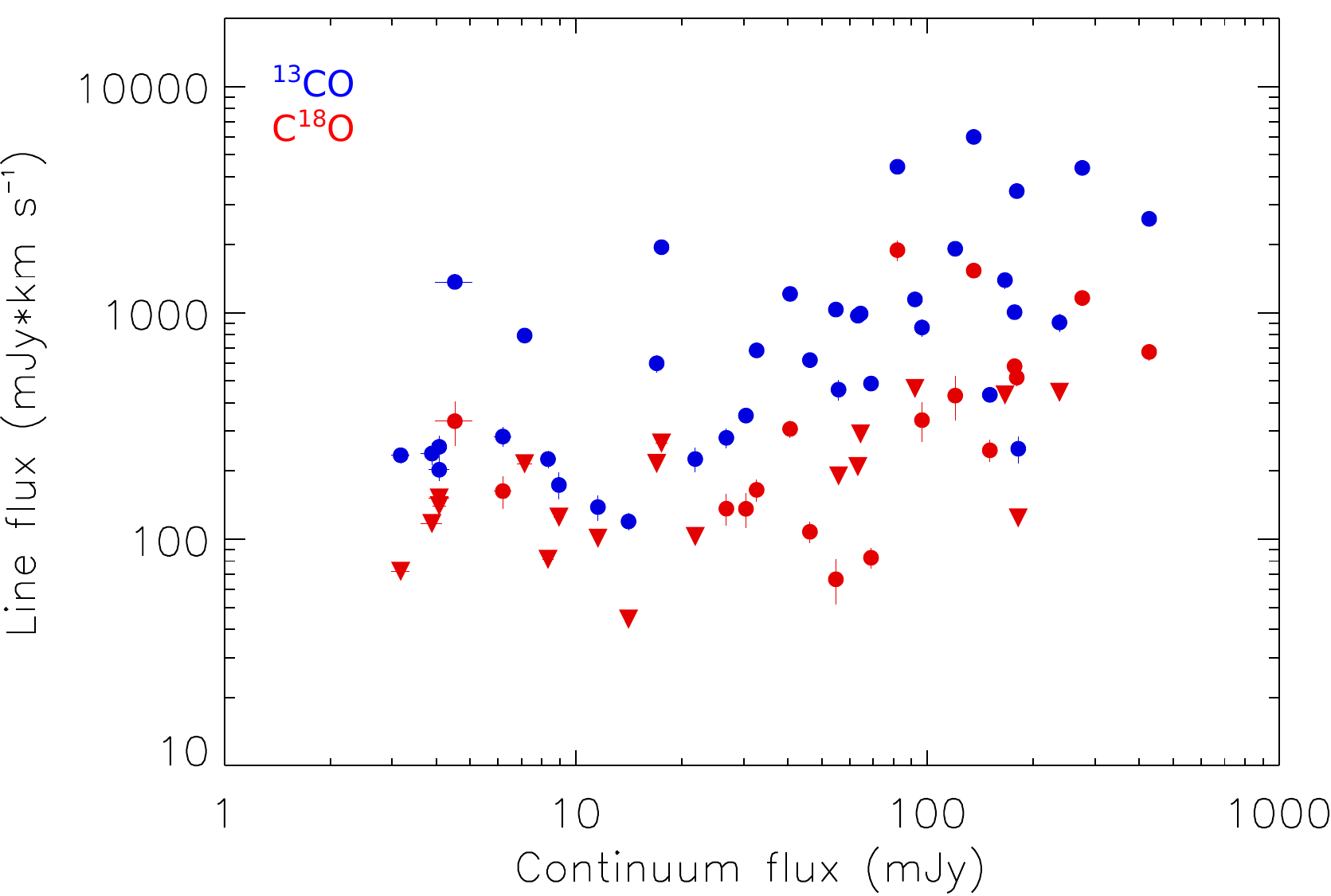}
\caption{Comparison of the integrated $^{13}$CO (blue) and C$^{18}$O (red) fluxes with the continuum fluxes of our sample disks. Triangles show the 4$\sigma$ upper limit for the C$^{18}$O non-detected disks.}\label{2flux}
\end{figure}

\begin{figure*}
\centering
\includegraphics[width=16.5cm]{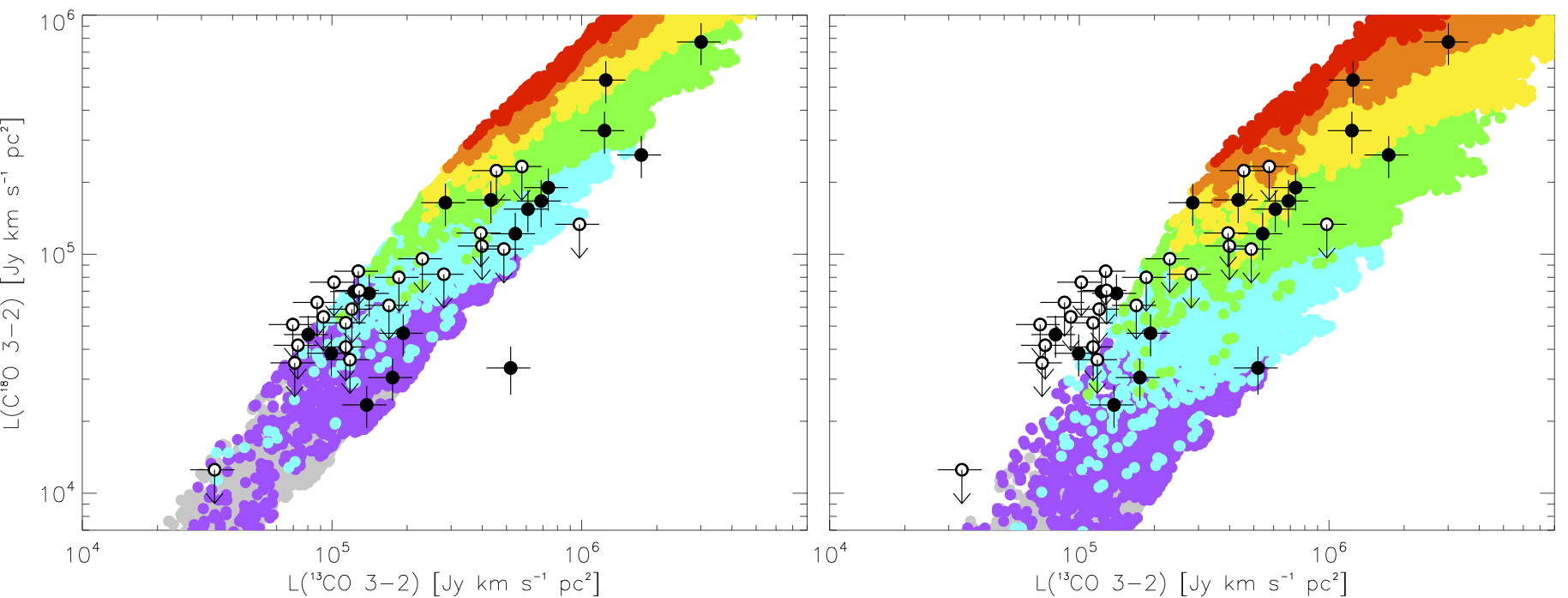}
\caption{$^{13}$CO (horizontal axis) and C$^{18}$O (vertical  axis) line luminosities from the physical-chemical models of protoplanetary disks (color dots) by \citet{Williams14} overlaid on the measured line luminosities from the velocity-aligned stacking method (data points with error bars). The C$^{18}$O non-detected disks are presented with open circles. Red, orange, yellow, green, blue, purple, and grey dots denote the models with disk masses of 10$^{-1}$, 3 $\times$ 10$^{-2}$, 10$^{-2}$, 3 $\times$ 10$^{-3}$, 10$^{-3}$, 3 $\times$ 10$^{-4}$, and 10$^{-4}$ $M_\sun$. The left panel is for the typical ISM abundance of C$^{18}$O, while the right panel is for the C$^{18}$O abundance that is three times lower the typical value.}\label{pltflux}
\end{figure*}

\begin{figure}
\centering
\includegraphics[width=8.5cm]{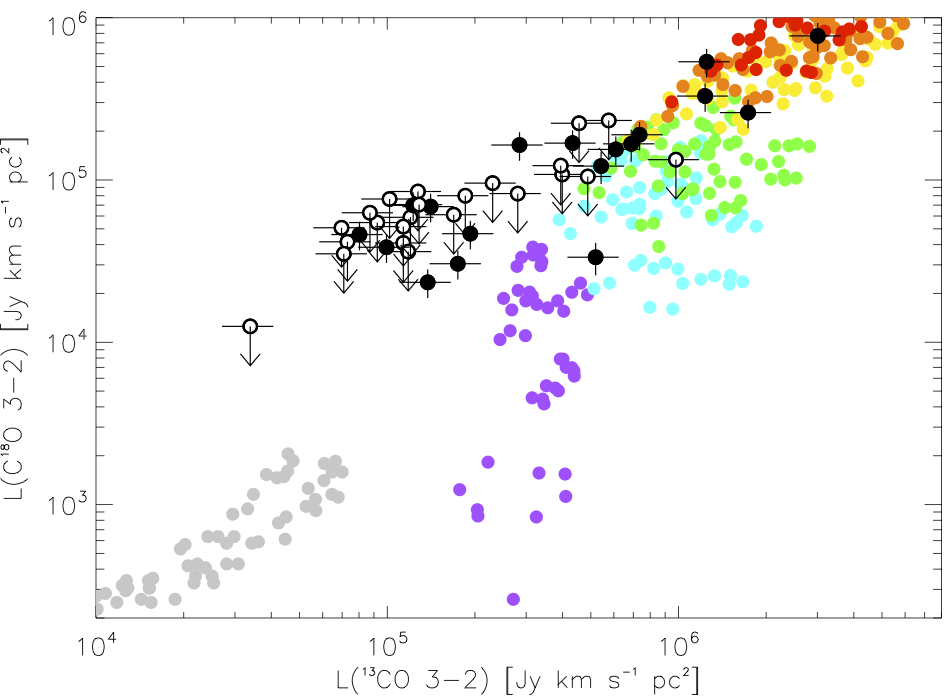}
\caption{$^{13}$CO (horizontal axis) and C$^{18}$O (vertical  axis) line luminosities from the physical-chemical models of protoplanetary disks (color dots) by \citet{Miotello16} overlaid on the measured line luminosities from the velocity-aligned stacking method (data points with error bars). The C$^{18}$O non-detected disks are presented with open circles. Red, orange, yellow, green, blue, purple, and grey dots denote the models with disk masses of 10$^{-1}$, 10$^{-2}$, 5 $\times$ 10$^{-3}$, 10$^{-3}$, 5 $\times$ 10$^{-4}$, and 10$^{-5}$, 10$^{-4}$ $M_\sun$.}\label{pltflux2}
\end{figure}

\begin{figure}
\centering
\includegraphics[width=7cm]{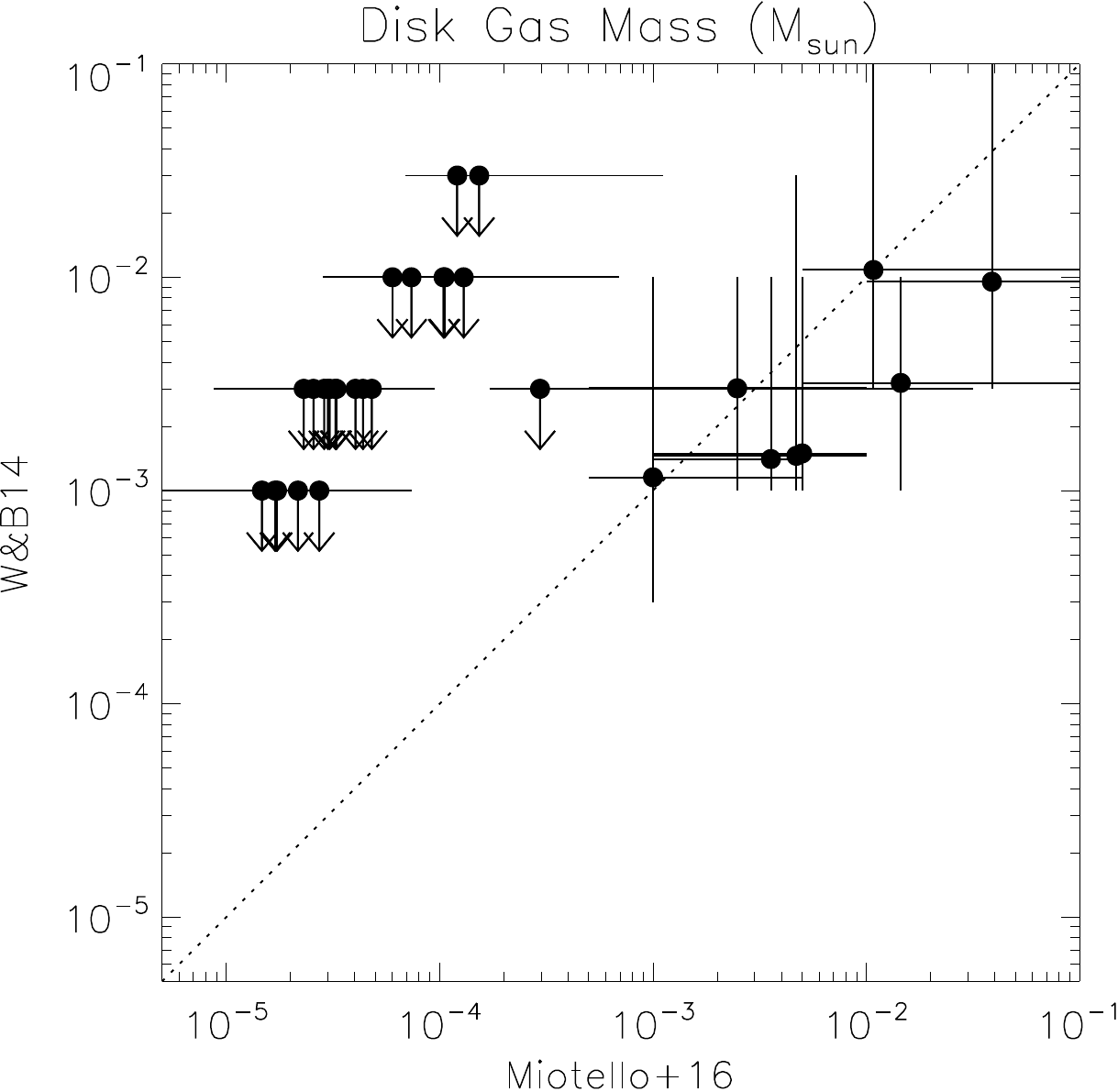}
\caption{Disk gas mass estimated from the $^{13}$CO and C$^{18}$O line luminosities with the disk models by \citet{Williams14} and \citet{Miotello16}, shown in the vertical and horizontal axes, respectively.}\label{mass2}
\end{figure}

Figure \ref{2flux} presents the comparison between the measured $^{13}$CO and C$^{18}$O line fluxes and the continuum fluxes. 
There is a clear correlation between the continuum and line fluxes.
Disks, which are bright in the continuum, tend to have higher line fluxes.
To estimate the disk gas mass from the $^{13}$CO and C$^{18}$O line fluxes, 
we followed the method in \citet{Ansdell16} and \citet{Miotello17}, 
and compared the measured $^{13}$CO and C$^{18}$O line luminosities with the grid of physical-chemical models of protoplanetary disks.

Figure \ref{pltflux} and \ref{pltflux2} present our measurements and the expected $^{13}$CO and C$^{18}$O line luminosities of protoplanetary disks with different masses from models by \citet{Williams14} and \citet{Miotello16}.
In \citet{Williams14}, two sets of models with different C$^{18}$O abundances were made, one with the typical ISM abundance (Fig.~\ref{pltflux} left) and one with a three times lower abundance to approximate selective photodissociation of C$^{18}$O (Fig.~\ref{pltflux} right).
For each set of models, there are a few measurements that do not match any model line luminosities. 
Nevertheless, when the two sets of models are combined, all the measured line luminosities can be reproduced with the models by \citet{Williams14}. 
On the other hand, \citet{Miotello16} modelled detailed disk chemistry and solved temperature structures with radiative transfer, 
but the temperature ranges covered by the grid of their models are not as wide as those in \citet{Williams14}. 
As discussed in \citet{Miotello16}, the temperature effect and spatial variation of the C$^{18}$O abundance are the primary difference between their work and \citet{Williams14}. 
As the number of models are smaller and the parameter ranges are narrower in \citet{Miotello16}, 
there are more measurements that cannot be reproduced with these models. 

We estimated disk masses with both models.
As discussed in Section \ref{method}, 
there can be a typical systematic uncertainty of 20\% in the measured integrated flux. 
Thus, in our estimate of disk mass, we included this 20\% uncertainty when the uncertainty of the measured integrated flux due to the noise is less than 20\%.
For disks detected in both $^{13}$CO and C$^{18}$O lines, 
we searched for the model line luminosities consistent with our measurements within the uncertainties.
The estimated disk mass is then adopted to be the mean mass of those model disks weighted by the difference between the model and observed line luminosities.
The uncertainty of the estimated disk mass is adopted to be the maximum and minimum disk mass of those models. 
6 out of the 18 disks detected in both the $^{13}$CO and C$^{18}$O lines can be reproduced with the models by \citet{Miotello16}. 
For the remaining 12 disks, we did not estimate their disk masses with the models by \citet{Miotello16}. 
For disks only detected in the $^{13}$CO line but not in the C$^{18}$O line, 
we placed the upper limit of disk mass with the models by \citet{Williams14}. 
We searched for the disk models with the $^{13}$CO line luminosities consistent with our measurements within the uncertainties and the C$^{18}$O line luminosities below our measured upper limit. 
Then, we adopted the maximum disk mass among these models to be the upper limit. 
In addition, we also estimated their disk masses with the fitting function of disk mass versus $^{13}$CO line luminosity from the grid of the disk models in \citet{Miotello17}. 
There are different fitting functions for face-on ($i=10\degr$) and edge-on ($i=80\degr$) disks. 
We have measured $i$ of all the detected disks. 
Therefore, the fitting function was selected for each disk based on whether $i$ is smaller or larger than 70$\degr$, the same as \citet{Miotello17}.
For a given line luminosity, the difference in the disk gas masses estimated with the two fitting functions for face-on and edge-on disks is $\sim$50\%.
In our sample, we found two disks, whose gas masses were estimated with the fitting function, that could suffer from this ambiguity in the disk inclination, \object{Sz~72} and \object{Sz~129}. 
Thus, their uncertainties in the disk gas mass are larger.

Figure \ref{mass2} compares the disk masses estimated with the models by \citet{Williams14} and \citet{Miotello16}. 
For disks which are detected in both lines and whose line luminosities can be reproduced with both models, 
the disk gas masses estimated from the two different models are consistent within the uncertainties, 
and their difference ranges from a factor of one to five.
For disks only detected in the $^{13}$CO line, with the models by \citet{Williams14}, we can only place a high upper limit of 10$^{-3}$ to 3 $\times$ 10$^{-2}$ $M_\sun$. 
In contrast, with the fitting function by \citet{Miotello17}, the disk masses of those disks are estimated to be on the order of 10$^{-5}$ to 10$^{-4}$ $M_\sun$.
All the disk masses estimated with the models by \citet{Miotello16} and \citet{Williams14} are listed in Table \ref{mdisk}.

\begin{table*}
\caption{Disk gas mass estimated from $^{13}$CO and C$^{18}$O line luminosities from the velocity-aligned stacking method for two model generations from \citet{Williams14} and \citet{Miotello16}.}\label{mdisk}
\centering
\begin{tabular}{lccccccc}
\hline\hline
 & \multicolumn{3}{c}{W\&B14} && \multicolumn{3}{c}{Miotello+16} \\
 \cline{2-4} \cline{6-8}
Source & Disk mass & Lower limit & Upper limit && Disk mass & Lower limit & Upper limit \\ 
& (10$^{-3}$ $M_\sun$) & (10$^{-3}$ $M_\sun$) & (10$^{-3}$ $M_\sun$) && (10$^{-3}$ $M_\sun$) & (10$^{-3}$ $M_\sun$) & (10$^{-3}$ $M_\sun$) \\
\hline \\
Sz~65 & ... & ... &  10 &&   0.074 &   0.037 &   0.205 \smallskip\\
J15450887$-$3417333 &   0.211 &   0.1 &   3 && ... & ... & ... \smallskip\\
Sz~68 &   0.552 &   0.3 &   3 && ... & ... & ... \smallskip\\
Sz~69 & ... & ... &   3 &&   0.044 &   0.019 &   0.089 \smallskip\\
Sz~71 & ... & ... &  10 &&   0.104 &   0.057 &   0.434 \smallskip\\
%Sz~72 & ... & ... &   1 &&   0.015 &   0.005 &   0.056 \smallskip\\
Sz~72 & ... & ... &   1 &&   0.015 &   0.002 &   0.056 \smallskip\\
Sz~73 &   0.234 &   0.1 &   1 && ... & ... & ... \smallskip\\
Sz~83 &   1.456 &   1 &  30 &&   4.698 &   1 &  10 \smallskip\\
Sz~84 &   0.368 &   0.1 &   3 && ... & ... & ... \smallskip\\
%Sz~129 & ... & ... &   1 &&   0.017 &   0.006 &   0.047 \smallskip\\
Sz~129 & ... & ... &   1 &&   0.017 &   0.006 &   0.073 \smallskip\\
RY~Lup &   3.197 &   1 &  10 &&  14.488 &   5 & 100 \smallskip\\
J16000236$-$4222145 &   1.154 &   0.3 &  10 &&   1 &   0.5 &   5 \smallskip\\
Sz~130 &   0.354 &   0.1 &   1 && ... & ... & ... \smallskip\\
MY~Lup &   2.682 &   1 &  30 && ... & ... & ... \smallskip\\
J16011549$-$4152351 &   9.534 &   3 & 100 &&  38.754 &  10 & 100 \smallskip\\
Sz~133 &   0.21 &   0.1 &   1 && ... & ... & ... \smallskip\\
Sz~88A & ... & ... &   1 &&   0.022 &   0.008 &   0.055 \smallskip\\
J16070384$-$3911113 &   1.406 &   1 &  10 &&   3.57 &   1 &   5 \smallskip\\
J16070854$-$3914075 & ... & ... &  30 &&   0.153 &   0.093 &   1.118 \smallskip\\
Sz~90 & ... & ... &   3 &&   0.029 &   0.011 &   0.066 \smallskip\\
Sz~95 & ... & ... &   3 &&   0.033 &   0.013 &   0.073 \smallskip\\
Sz~96 & ... & ... &   3 &&   0.026 &   0.01 &   0.061 \smallskip\\
J16081497$-$3857145 & ... & ... &   3 &&   0.04 &   0.016 &   0.093 \smallskip\\
Sz~98 & ... & ... &  30 &&   0.121 &   0.069 &   0.597 \smallskip\\
Sz~100 &   0.41 &   0.3 &   1 && ... & ... & ... \smallskip\\
Sz~103 & ... & ... &   1 &&   0.017 &   0.006 &   0.047 \smallskip\\
J16083070$-$3828268 &  10.821 &   3 & 100 &&  10.79 &   5 & 100 \smallskip\\
V856~Sco & ... & ... &  10 &&   0.06 &   0.028 &   0.132 \smallskip\\
Sz~108B &   0.552 &   0.3 &   3 && ... & ... & ... \smallskip\\
J16085373$-$3914367 & ... & ... &   3 &&   0.03 &   0.012 &   0.068 \smallskip\\
Sz~111 &   3.017 &   1 &  10 &&   2.48 &   0.5 &  10 \smallskip\\
J16090141$-$3925119 & ... & ... &   3 &&   0.295 &   0.171 &  31.579 \smallskip\\
Sz~114 &   1.7 &   1 &  30 && ... & ... & ... \smallskip\\
J16092697$-$3836269 & ... & ... &   3 &&   0.031 &   0.012 &   0.069 \smallskip\\
J160934.2$-$391513 & ... & ... &   3 &&   0.033 &   0.013 &   0.072 \smallskip\\
J16093928$-$3904316 & ... & ... &   1 &&   0.027 &   0.01 &   0.074 \smallskip\\
Sz~118 & ... & ... &  10 &&   0.13 &   0.075 &   0.695 \smallskip\\
J16100133$-$3906449 & ... & ... &   3 &&   0.048 &   0.021 &   0.095 \smallskip\\
J16101984$-$3836065 & ... & ... &   3 &&   0.023 &   0.009 &   0.057 \smallskip\\
J16102955$-$3922144 & ... & ... &  10 &&   0.106 &   0.058 &   0.445 \smallskip\\
Sz~123A &   1.494 &   1 &  10 &&   5 &   1 &  10 \smallskip\\
\hline
\end{tabular}
%\tablefoot{}
%\tablebib{(1) \citet{branch83}; (2) \citet{phillips87}; (3) \citet{barbon90}}
\end{table*}

\section{Discussion}

\subsection{Spectroscopic and Dynamical Stellar Mass}\label{mstar}

\begin{figure*}
\centering
\includegraphics[width=14cm]{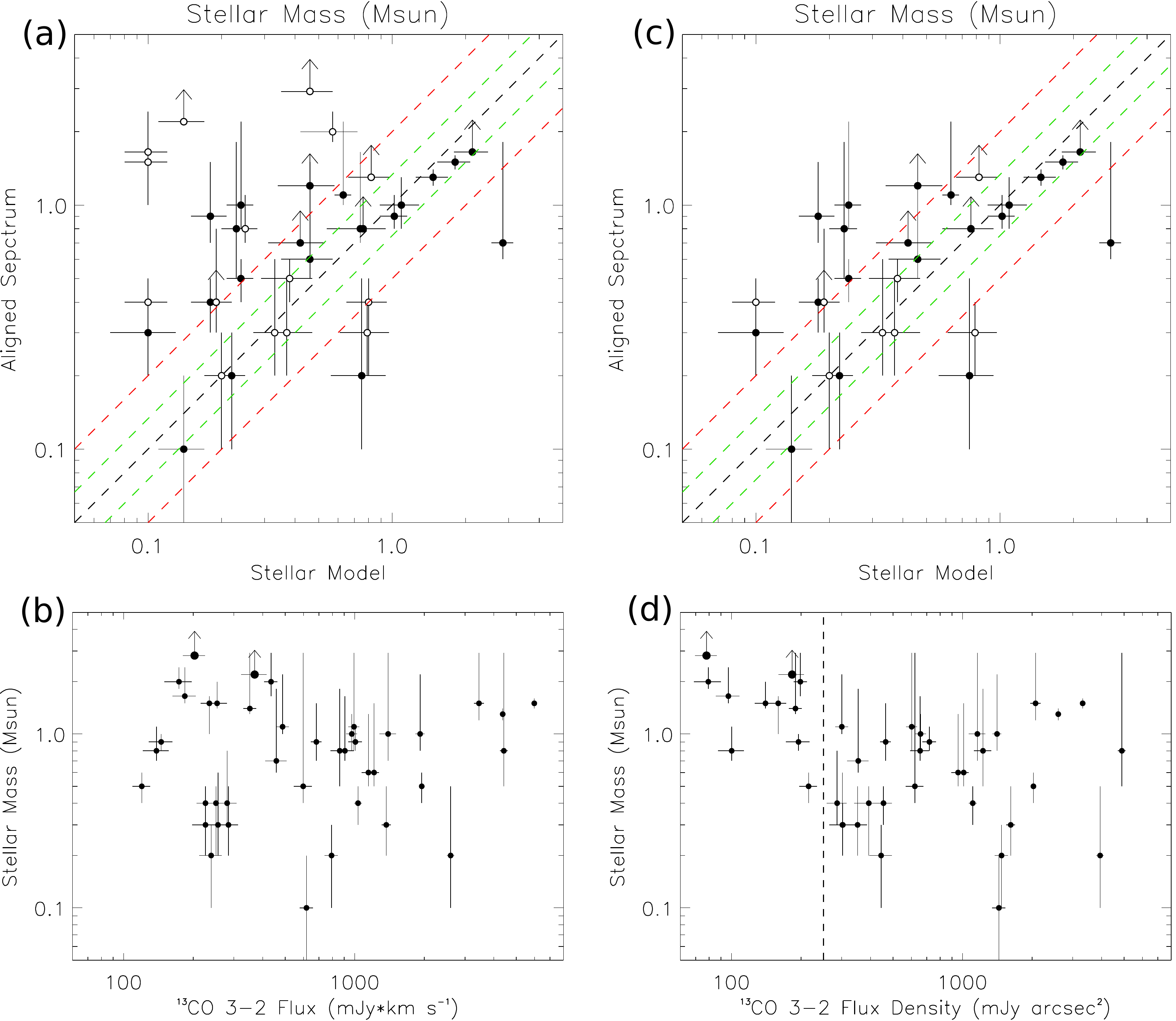}
\caption{(a) Stellar masses estimated with the spectroscopic method (horizontal axis; \citet{Alcala14,Alcala17}) and with the dynamical information traced by the $^{13}$CO emission from this work (vertical axis). Filled and open symbols present the disks with the integrated $^{13}$CO fluxes above and below 400 mJy km s$^{-1}$, respectively. Black, green, and red dashed lines denote the regions where the differences in mass are 0, 25\%, and 50\%, respectively. (b) Stellar masses estimated with the dynamical information from this work (vertical axis) compared with measured integrated $^{13}$CO fluxes (horizontal axis). (c) Same as (a) but with a subsample of lowest $^{13}$CO flux density discarded. (d) Same as (b) but the horizontal axis presents $^{13}$CO flux density. A vertical dashed line in (d) denotes our selection criterion for the subsample shown in (c).}\label{corrms}
\end{figure*}

We estimated the stellar masses of the 41 YSOs from their gas kinematics traced by the $^{13}$CO emission with the velocity-aligned stacking method. 
The stellar masses of 37 of them have also been estimated by comparing their stellar effective temperatures and luminosities with the evolutionary models by \citet{Siess00} as discussed by \citet{Alcala14, Alcala17}.
Figure \ref{corrms}a compares the stellar masses estimated with the dynamical information from this work and the spectroscopic information in the literature. 
There is a group of sources showing a correlation between dynamically and spectroscopically determined stellar masses. 
In 21 out of 37 ($\sim$60\%) sources, the two masses are consistent within the 2$\sigma$ uncertainties, or their difference is less than 50\%. 
We note that there are several sources having their dynamical stellar masses significantly larger than their spectroscopic stellar masses by a factor of several (upper left corner in Fig.~\ref{corrms}a).
We compare the estimated stellar masses as a function of integrated $^{13}$CO flux (Fig.~\ref{corrms}b). 
We find that high estimated stellar masses tend to be associated with lower integrated $^{13}$CO fluxes. 
S/N ratios per unit area and unit velocity (i.e., flux density) are proportional to integrated flux over disk size and line width. 
The projected disk area on the plane of the sky is proportional to $\cos i$, 
and the line width of disk rotation is proportional to $\sin i\sqrt{M_\star}$.
Thus, we compute flux densities of the disks as $F_{\rm ^{13}CO} / (\cos i \sin i\sqrt{M_\star}$) and compare with our estimated stellar masses in Fig.~\ref{corrms}d. 
A sub-group with high estimated stellar mass and low flux density is clearly seen. 
Thus, their stellar masses can be overestimated due to their low fluxes, as discussed in Section \ref{demo2}.
In Fig.~\ref{corrms}c, we exclude that sub-group with flux density less than 250 mJy arcsec$^{-2}$.  
As a result, the correlation between our dynamically estimated stellar masses and the spectroscopically determined stellar masses becomes clear. 

\begin{figure}
\centering
\includegraphics[width=7cm]{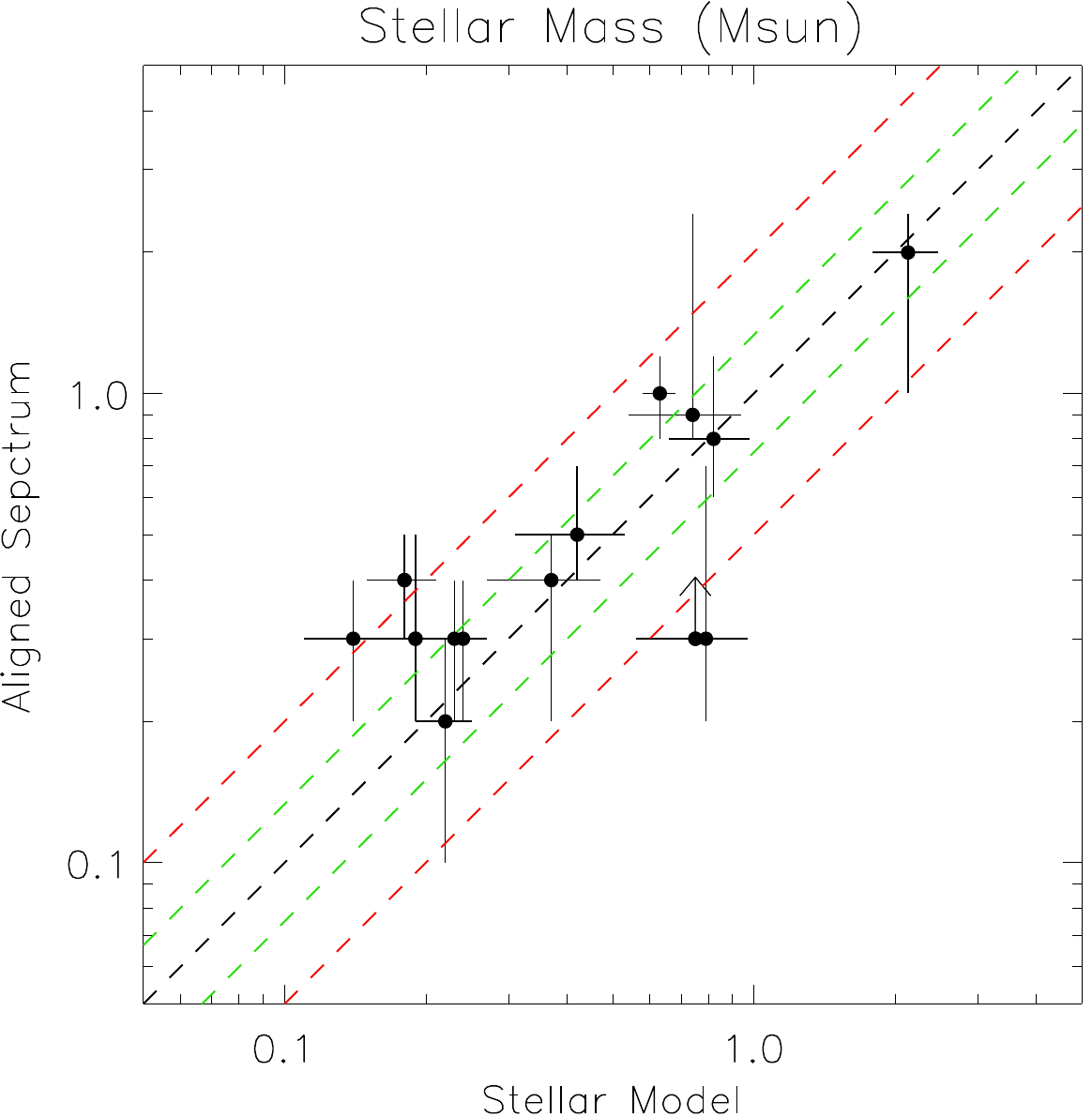}
\caption{Same as Fig.~\ref{corrms}a but fixed PA and $i$ were adopted from the continuum results \citep{Ansdell16,Tazzari17} when we estimated the stellar masses with the $^{13}$CO emission.}\label{corrms2}
\end{figure}

The main uncertainty in our estimated stellar mass is from the constraint on the inclination angle. 
As discussed in Section \ref{method}, it requires more spatial information to better constrain an inclination angle of a disk. 
The inclination angles of a subsample of the disks in this Lupus survey have also been measured from the continuum emission (Fig.~\ref{corrdisk}), 
which has much higher S/N ratios to constrain disk orientations as compared to the $^{13}$CO data. 
Thus, we have also estimated the stellar masses of this subsample with our method but adopted PA and $i$ from the continuum results \citep{Ansdell16, Tazzari17}. 
Hence, there are only two free parameters, $M_\star$ and $V_{\rm sys}$, in this analysis.  
The re-estimated stellar masses are typically consistent with the original values within the 1$\sigma$ to 2$\sigma$ uncertainties because our method can trace the disk orientation reasonably well (Fig.~\ref{corrdisk}), 
and they tend to become closer to the spectroscopic stellar masses. 
The estimated stellar masses with the fixed PA and $i$ are presented and compared with their spectroscopic stellar masses in Table \ref{mstab2} and Fig.~\ref{corrms2}.
In this subsample, the dynamical stellar masses are in a very good agreement with the spectroscopic stellar masses. 
The dynamical and spectroscopic stellar masses in 10 out of 16 sources are consistent within the 1$\sigma$ uncertainties, 
and they are consistent within 2$\sigma$ in all the sources.
The mean difference between the dynamical and spectroscopic stellar masses in this subsample is 0.15 $M_\sun$.

Our results are similar to a recent report by \citet{Simon17}.  
They analyzed and modelled Keplerian rotation of disks around 25 pre-main sequence stars observed with ALMA at angular resolutions of 0\farcs2 to 0\farcs8.
With the high S/N-ratio data and detailed modelling, the achieved 1$\sigma$ uncertainties in dynamical stellar mass in their studies are less than 0.1 $M_\sun$, 
and the mean uncertainty is 0.03 $M_\sun$.
They found that in 7 out of 25 (30\%) pre-main sequence stars, their dynamical stellar masses range from 1.0 $M_\sun$ to 2.3 $M_\sun$, and their luminosities ($\log L = -1.57\mbox{--}0.37$ $L_\sun$) and effective temperatures ($\log T_{\rm eff} = 3.51\mbox{--}3.64$ K) are inconsistent with the expectation for stars with such masses from the stellar evolutionary models. 
They suggest that these sources are unresolved binary or multiple systems. 
For the remaining stars with masses ranging from 0.1 $M_\sun$ to 1.1 $M_\sun$, their estimated dynamical masses are consistent with their spectroscopic masses within the uncertainties. 
Although the accuracy of our estimated stellar mass is not sufficiently high to calibrate stellar evolution models due to the low S/N ratios of the data, 
the observed trend between the dynamical and spectroscopic stellar masses of 30 sources in Fig.~\ref{corrms}c demonstrates the robustness of the velocity-aligned stacking method to estimate stellar masses, 
and our estimated dynamical masses of the subsample of 16 sources, where the inclination angles are better constrained, are in a good agreement with the stellar evolutionary models of the mass range from 0.1 $M_\sun$ to 2 $M_\sun$. 

\begin{table*}
\caption{Subsample of dynamically and spectroscopically determined stellar masses}\label{mstab2}
\centering
\begin{tabular}{lcccc}
\hline\hline
Source & Dynamical $M_\star$ & $V_{\rm sys}$ & Spectroscopic $M_\star$ & Difference \\ 
& ($M_\sun$) & (km s$^{-1}$) & ($M_\sun$) & ($\sigma$)\\
\hline \\
J15450887$-$3417333 & 0.3$\pm$0.1 & 4.8$^{+0.3}_{-0.3}$ & 0.14$\pm$0.03 & 1.5 \smallskip\\
Sz~68 & 2.0$^{+0.4}_{-1.0}$ & 5.2$^{+0.1}_{-0.4}$ & 2.13$\pm$0.34 & 0.2 \smallskip\\
Sz~69 & 0.3$\pm$0.1& 5.5$^{+0.1}_{-0.4}$ & 0.19$\pm$0.03 & 1.1 \smallskip\\
Sz~71 & 0.5$^{+0.2}_{-0.1}$ & 3.6$^{+0.1}_{-0.2}$ & 0.42$\pm$0.11 & 0.5 \smallskip\\
Sz~73 & 0.8$^{+0.4}_{-0.2}$ & 4.3$^{+0.5}_{-0.3}$ & 0.82$\pm$0.16 & 0.1 \smallskip\\
Sz~83 & >0.3 & 4.7$^{+0.2}_{-0.3}$ & 0.75$\pm$0.19 & ... \smallskip\\
Sz~84 & 0.4$\pm$0.1 & 5.2$^{+0.3}_{-0.2}$ & 0.18$\pm$0.03 & 2.1 \smallskip\\
J16000236$-$4222145 & 0.3$\pm$0.1 & 4.1$\pm$0.2 & 0.24$\pm$0.03 & 0.6 \smallskip\\
Sz~130 & 0.4$^{+0.1}_{-0.2}$ & 4.6$^{+0.1}_{-0.4}$ & 0.37$\pm$0.10 & 0.1 \smallskip\\
Sz~133 & 1.0$\pm$0.2 & 5.3$^{+0.2}_{-0.5}$ & 0.63$\pm$0.05 & 1.8 \smallskip\\
Sz~90 & 0.3$^{+0.4}_{-0.1}$ & 5.5$^{+0.1}_{-0.4}$ & 0.79$\pm$0.18 & 1.1 \smallskip\\
Sz~98 & 0.9$^{+1.5}_{-0.1}$ & 3.2$^{+0.5}_{-0.4}$ & 0.74$\pm$0.20 & 0.7 \smallskip\\
Sz~100 & 0.4$\pm$0.1 & 1.9$\pm$0.2 & 0.18$\pm$0.03 & 2.1 \smallskip\\
Sz~108B & 0.3$^{+0.2}_{-0.1}$ & 2.5$^{+0.2}_{-0.1}$ & 0.19$\pm$0.03 & 1.1 \smallskip\\
Sz~114 & 0.3$\pm$0.1 & 5.0$^{+0.1}_{-0.2}$ & 0.23$\pm$0.03 & 0.7 \smallskip\\
J16102955$-$3922144 & 0.2$\pm$0.1 & 3.5$^{+0.3}_{-0.2}$ & 0.22$\pm$0.03 & 0.2 \smallskip\\
\hline
\end{tabular}
\tablefoot{In this subsample, the dynamical $M_\star$ is estimated with our method described in Section \ref{method} but with fixed PA and $i$ adopted from the continuum analysis by \citet{Ansdell16} and \citet{Tazzari17}. The spectroscopic  $M_\star$ is obtained from \citet{Alcala14,Alcala17}, which is estimated with the stellar evolutionary model by \citet{Siess00}.}
%\tablebib{(1) \citet{branch83}; (2) \citet{phillips87}; (3) \citet{barbon90}}
\end{table*}

\subsection{Gas-to-Dust Ratio and Disk Mass}

\begin{figure}
\centering
\includegraphics[width=9cm]{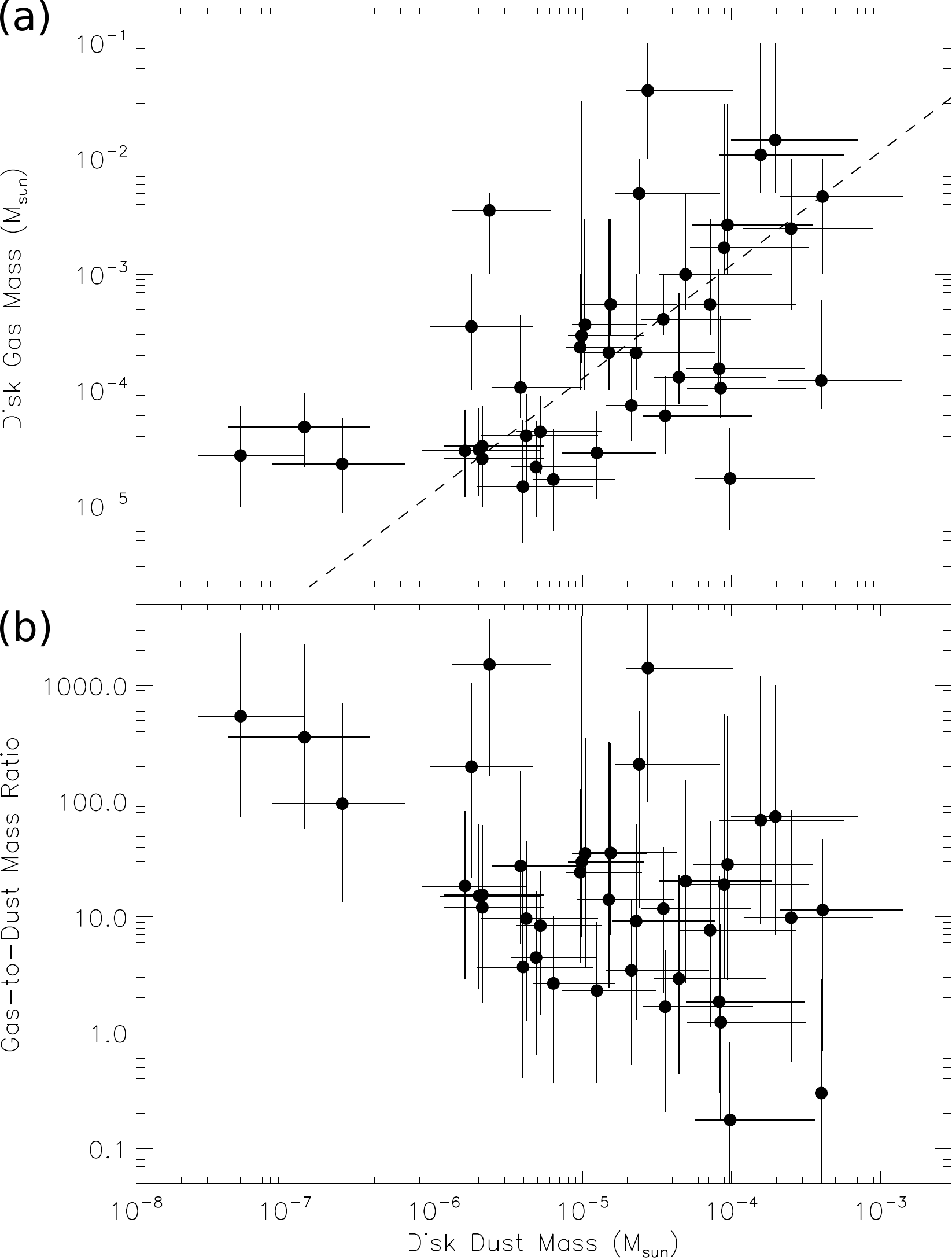}
\caption{(a) Disk gas mass as a function of disk dust mass. (b) Gas-to-dust mass ratio as a function of disk dust mass. The disk dust masses are estimated with the fitting function in \citet{Miotello17}. A dashed line presents the best linear fit. Only the disks with their dust masses larger than 10$^{-6}$ $M_\sun$ were included in the linear fit.}\label{g2d}
\end{figure}

Our analysis enabled more measurements of disk gas masses from $^{13}$CO and C$^{18}$O data, 
especially for disks with an integrated $^{13}$CO flux lower than 500 mJy km s$^{-1}$, as compared to  \citet{Ansdell16}. 
Figure \ref{g2d} presents our disk gas mass estimated with the $^{13}$CO and C$^{18}$O emission and gas-to-dust mass ratios compared with the disk dust mass. 
In Fig.~\ref{g2d}, the gas mass of the disks detected in the $^{13}$CO line but not in the C$^{18}$O line is estimated with the fitting function in \citet{Miotello17}.   
For the disks detected in both the $^{13}$CO and C$^{18}$O lines, 
their disk gas masses are estimated with the grid of the disk models by \citet{Miotello16} when their line luminosities can be reproduced with these disk models. 
Otherwise their disk gas masses are estimated with the grid of the disk models by \citet{Williams14}.
As shown in Fig.~\ref{mass2}, the estimated gas masses with the two models are consistent within the uncertainties, when the observed line luminosities can be reproduced by both models.
The disk dust mass shown in Fig.~\ref{g2d} is estimated with the continuum flux in \citet{Ansdell16} and the fitting function in \citet{Miotello17}. 
We note that there are a few disks having estimated gas masses of a few $\times$ 10$^{-5}$ $M_\sun$ but very low dust masses on the order of 10$^{-7}$ $M_\sun$, 
and all their derived gas-to-dust mass ratios are a factor of several hundred. 
On the other hand, the distribution of the gas-to-dust ratios of the disks with their dust masses higher than 10$^{-6}$ $M_\sun$ is relatively flat and has a median of 12. 
When this median gas-to-dust mass ratio is applied to disks with very low dust masses, assuming that the gas-to-dust mass ratio does not strongly depend on disk dust mass (as shown below), 
we expect their disk gas masses traced by the $^{13}$CO emission to be on the order of 10$^{-6}$ $M_\sun$. 
That is ten times lower than the minimum disk mass of the grid of the disk models by \citet{Miotello16}, which is 10$^{-5}$ $M_\sun$. 
In addition, Fig.~4 in \citet{Miotello17} shows that the gas mass of a disk with a $^{13}$CO line luminosity lower than 7 $\times$ 10$^4$ Jy km s$^{-1}$ pc$^2$ can be potentially lower than 10$^{-5}$ $M_\sun$, but that low mass regime was not explored. 
Thus, we consider that the gas mass estimates of those disks with dust masses lower than 10$^{-6}$ $M_\sun$ may not be reliable, and we exclude those disks in the following discussion. 

Figure \ref{g2d}a shows that high disk gas masses estimated with the $^{13}$CO and C$^{18}$O emission tend to be associated with high disk dust masses estimated with the continuum.  
We performed a linear fit to the logarithms of the disk gas and dust masses by using the routine of a Bayesian approach with errors of both axes written by \citet{Kelly07}, {\it linmix\_err}. 
The error bars of the disk masses are estimated from the mass ranges of the grid of the disk models showing consistent line and continuum luminosities with the observations within the uncertainties \citep[Fig.~1 and 4 in][]{Miotello17}, 
so they are not conventional 1$\sigma$ uncertainties with normal distributions.  
This is different from the assumption in {\it linmix\_err}, which considers errors are normally distributed.
For simplicity and to be conservative, we adopted the logarithms of the ratios of the upper limits to the measurements as 1$\sigma$ uncertainties in the linear fit, 
and thus the uncertainty in our linear fit is unlikely underestimated.
The fitting result shows that the logarithms of the disk gas and dust masses are possibly correlated with a slope of 1.1$\pm$0.5 and a correlation coefficient of 0.8$\pm$0.2 in the logarithmic scale. 
As a consequence, the logarithms of the gas-to-dust mass ratio and the dust mass do not have any significant correlation but with the possible flat value of 12, despite a large scatter. 
We did not include the disks which are not detected in the $^{13}$CO emission in the linear fit because it is not straightforward to estimate the upper limits of their line 
fluxes and their resulting disk gas masses without knowing their line widths, systemic velocities, and disk sizes. 
Nevertheless, as demonstrated in Section \ref{method}, we are able to detect disks with an integrated flux of a few hundred mJy km s$^{-1}$ at a distance of 150 pc, depending on the actual disk parameters. 
Thus, the upper limits of their $^{13}$CO line luminosities are most likely on the order of 10$^5$ Jy km s$^{-1}$ pc$^2$, corresponding to upper limits of their disk gas masses of a few $\times$ 10$^{-5}$ $M_\sun$ with the fitting function in \citet{Miotello17}.
On the other hand, the disk dust masses of the disks undetected in $^{13}$CO are mostly less than 10$^{-5}$ $M_\sun$. 
Only one undetected disk has a dust mass of more than 10$^{-4}$ $M_\sun$, and three have a mass of more than 10$^{-5}$ $M_\sun$. 
Therefore, when we include these disks and adopt an upper limit of the disk gas mass of 10$^{-4}$ $M_\sun$ in the linear fit$\footnotemark$, 
our results remain unchanged. 
We have also verified that our results do not depend on whether the disk dust masses are adopted from \citet{Ansdell16} or from \citet{Miotello17},  
and that the results are not affected by including an additional 10\% uncertainty due to the absolute flux calibration in the analysis.

\footnotetext{When non-detections are included in the analysis, {\it linmix\_err} introduces an indicator variable for detection and non-detection in the likelihood function to account for upper limits provided by these non-detections.}

However, we note that the scatter in Fig.~\ref{g2d} is large with a 1$\sigma$ intrinsic scatter of 0.2$\pm$0.2 estimated from {\it linmix\_err}.  
The estimated gas masses and the derived gas-to-dust mass ratios span over three to four orders of magnitude at a given bin of the disk dust masses. 
We subtracted the derived correlation from the distribution of the disk gas and dust masses and measured the standard deviation of the residual gas masses to be 0.9 dex. 
Such a large scatter and wide range of gas-to-dust mass ratios likely suggests that additional important physics and chemistry are not yet captured in the current disk mass estimates, leading to the inaccuracy in the estimated disk gas masses. 
As discussed in \citet{Miotello17}, 
the estimates of the disk gas masses with the models by \citet{Williams14} and \citet{Miotello16} are based on the assumption that there is no depletion of volatile carbon and oxygen, 
but observations have shown hints of carbon depletion in protoplanetary disks \citep{Schwarz16, McClure16}, 
which can lead to inaccurate estimates of the disk gas masses. 

In addition, we note that although the line fluxes are correlated with the continuum fluxes (Fig.~\ref{2flux}), the presence of the possible correlation between the disk gas and dust masses is subject to the disk models adopted to estimate the gas masses from the line fluxes.
The possible correlation between the disk gas and dust masses is only present when we include the data points of the disk gas masses estimated with the fitting function in \citet{Miotello17} for those disks detected in the $^{13}$CO line but not in the C$^{18}$O line and having gas masses around a few $\times$ 10$^{-5}$ $M_\sun$. 
For these disks undetected in the C$^{18}$O line, 
when we estimate their disk gas masses with the disk models by \citet{Williams14}, 
we can only obtain upper limits of 10$^{-3}$ $M_\sun$ or above (Fig.~\ref{mass2}). 
%Thus, to avoid a possible bias (if any) in the gas mass estimate for the disks without C$^{18}$O detection, 
%we have also performed the linear fit by adopting the upper limits of the gas masses estimated with the disk models by \citet{Williams14} instead of adopting the gas masses estimated with the fitting function in \citet{Miotello17}. 
When we performed the linear fit by adopting the upper limits of the gas masses estimated with the disk models by \citet{Williams14} instead of adopting the gas masses estimated with the fitting function in \citet{Miotello17} or by excluding this subsample,  
we found that the correlation between disk gas and dust masses becomes insignificant. 
Therefore, measuring gas masses of disks detected in $^{13}$CO but not in C$^{18}$O is essential to constrain the possible correlation between the disk gas and dust masses. 
Deeper observations to detect their C$^{18}$O emission are needed to compare with the disk models and examine the current gas mass estimate. 

\begin{figure}
\centering
\includegraphics[width=9cm]{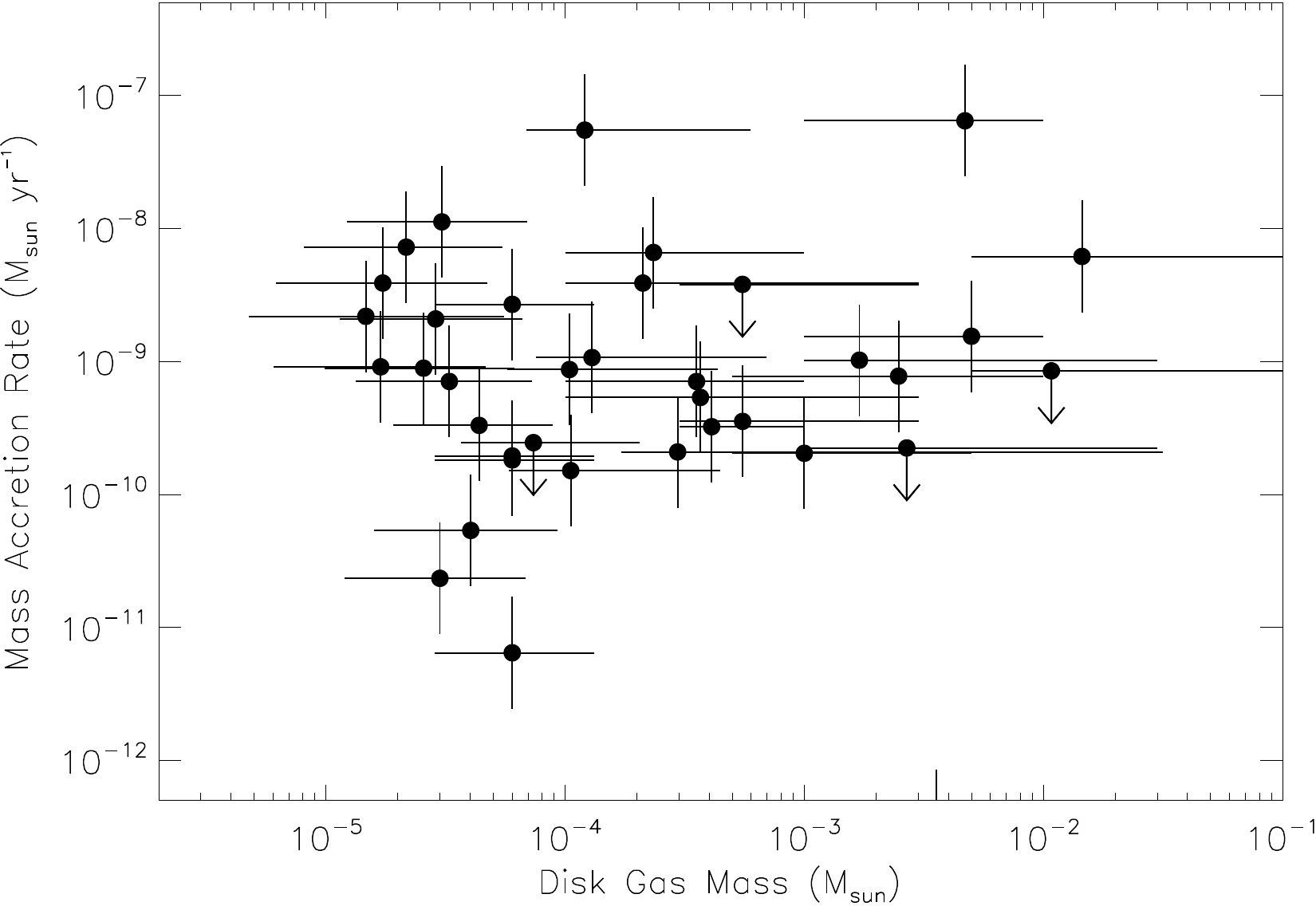}
\caption{Mass accretion rate as a function of disk gas mass. The mass accretion rates are obtained from \citet{Alcala14, Alcala17}.}\label{macc}
\end{figure}

Correlations between mass accretion rates and disk dust masses have been observed in this Lupus sample \citep{Manara16} and in the Chamaeleon~I sample \citep{Mulders17}.  
The presence of a correlation between mass accretion rates and disk total masses is expected in the theory of viscously evolving disks \citep[e.g.,][]{Hartmann98, Jones12, Rosotti17, Lodato17}.  
However, no significant correlation between the mass accretion rate and the disk gas mass estimated from the $^{13}$CO and C$^{18}$O lines was found in these studies. 
Here, with more measurements of the disk gas mass, 
we revisit the relation between the mass accretion rate and the disk gas mass (Fig.~\ref{macc}). 
The mass accretion rates are obtained from \citet{Alcala14,Alcala17}.
In \citet{Manara16}, several disks only had upper limits of gas masses ranging from 5 $\times$ 10$^{-4}$ to 10$^{-3}$ $M_\sun$. 
Here, their gas masses are re-estimated to be on the order of 10$^{-5}$ $M_\sun$ with the fitting function in \citet{Miotello17}. 
However, after adding these new measurements in the analysis, 
we still find no significant correlation between the mass accretion rate and the disk gas mass, 
regardless of including the data points with only the upper limits of the disk gas mass or the mass accretion rate, the same as the results in \citet{Manara16} and \citet{Miotello17},
even though high disk gas masses tend to be associated with high disk dust masses (Fig.~\ref{g2d}a). 
We found that even if the disk gas masses are correlated with the disk dust masses, 
still no correlation between the mass accretion rate and the disk gas mass is expected because of the large uncertainties in deriving the total disk gas masses from the $^{13}$CO and C$^{18}$O emission (see \cite{Miotello17} for a detailed discussion).
We have generated simulated samples having random dust masses ranging from 10$^{-3}$--10$^{-6}$ $M_\sun$, 
and we computed their disk gas masses and mass accretion rates with the observed correlations and randomly added the observed scatters \citep[this work and][]{Mulders17}. 
Then we compared the simulated disk gas masses and mass accretion rates, and indeed did not find any significant correlation.
With the current data, disks with gas masses less than 10$^{-5}$ $M_\sun$ cannot be probed. 
To investigate the correlation between the disk gas mass and the mass accretion rate with the current uncertainties in deriving the disk gas mass, measurements of disk gas masses at this low-mass end ($<$10$^{-5}$ $M_\sun$) to expand the range of disk masses in the sample are needed.   

\section{Summary}
In order to extract more information from molecular-line data of the ALMA surveys on protoplanetary disks and to study disk properties in molecular lines, 
we re-analyzed the ALMA $^{13}$CO (3--2) and C$^{18}$O (3--2) data of 88 Lupus YSOs with the velocity-aligned stacking method. 
With this method, we search for detection and estimate the dynamical stellar masses of the YSOs and the position angle of the major axis, inclination angle, and systemic velocity of their associated disks with the $^{13}$CO data by maximizing the auto-correlation between the data and the various generated Keplerian rotational patterns. 
After applying this method, the S/N ratios of these data are enhanced, 
and we obtain $^{13}$CO detections in 41 disks and C$^{18}$O detections in 18 disks. 
11 of them were not detected in the $^{13}$CO emission in the previous study, and there are six disks detected in the $^{13}$CO emission in the previous study but not in our analysis \citep{Ansdell16}. 
Our main results are summarized below.
\begin{enumerate}
\item{We have tested the velocity-aligned stacking method to estimate stellar mass and disk orientation with observational and synthetic data. 
We measured the disk orientations of \object{RY~Lup}, \object{J16083070$-$3828268}, and \object{Sz~83} from their moment 0 maps of the $^{13}$CO emission and their stellar masses and systemic velocities from the velocity structures in the PV diagrams of the $^{13}$CO emission. 
The estimates from our velocity-aligned stacking method are consistent with these measurements from the direct images within the uncertainties. 
Our test with synthetic data shows that with these ALMA observations, this method can detect disks and obtain robust measurements when the integrated line flux is $\gtrsim$400 mJy km s$^{-1}$. 
When the integrated line flux is as low as 400 mJy km s$^{-1}$, the stellar mass can be overestimated because of the limitation in constraining disk orientation. 
Nevertheless, the measurement of the integrated line flux remains robust. 
The uncertainty in the measured flux due to the limited sensitivity with this method is typically 20\% and exceptionally it can be a factor of two.}

\item{The position angles of the disks estimated from our method are clearly correlated with those from the continuum emission, except for two outliers where the difference is more than 45\degr. 
The inclination angles of the disks estimated from our method and from the continuum emission are also correlated but have a larger scatter, as the uncertainties in the inclination angles are larger. 
Therefore, our method can detect independently and consistently the disk orientations, and hence, also reveal a disk orientation in the absence of a clear direct detection in continuum or lines. 
Our measured fluxes are consistent with the measurements in \citet{Ansdell16} within a few percent to 30\% for bright disks with integrated fluxes larger than 1000 mJy km s$^{-1}$. The difference in the measured fluxes ranges from 10\% to a factor of five for fainter disks. 
Compared to \citet{Ansdell16}, our method additionally detects several more disks with integrated fluxes of 100--500 mJy km s$^{-1}$. }

\item{The stellar masses of 37 of our $^{13}$CO detected disks have been spectroscopically determined with stellar evolutionary models. 
After excluding a subsample of disks with a low $^{13}$CO flux density, where their dynamical stellar masses can be overestimated with our method, 
a correlation between the dynamical and spectroscopic stellar masses of 30 sources is seen, demonstrating the robustness of the velocity-aligned stacking method to estimate stellar masses. 
In a subsample of 16 sources, where the inclination angles are better constrained with the continuum data, we have also estimated their stellar masses with our method but adopted PA and $i$ from the continuum results. 
In these sources, our estimated dynamical masses are consistent with their spectroscopic masses within the uncertainties with a mean difference between the two masses of 0.15 $M_\sun$. 
Thus, the dynamical masses in this subsample are in a good agreement with the stellar evolutionary models of the mass range from 0.1 $M_\sun$ to 2 $M_\sun$.}

\item{We estimate the total gas masses of our detected disks by comparing the measured $^{13}$CO and C$^{18}$O line luminosities with the grid of the disk models by \citet{Williams14} and \citet{Miotello16}.
We find that high disk gas masses estimated with the $^{13}$CO and C$^{18}$O emission tend to be associated with high disk dust masses estimated with the continuum, hinting at a linear relation, and thus, the derived gas-to-dust ratios show no dependence on the disk dust mass. 
However, we note that the distribution of the disk gas and dust masses has a large scatter, 
and the derived gas-to-dust mass ratios span over three to four orders of magnitude at a given bin of the disk dust mass, likely suggesting that additional important physics and chemistry are not yet captured in the current disk mass estimates, leading to the inaccuracy in the estimated disk gas masses. 
%implying inaccuracy in our current estimates of the disk gas mass from the $^{13}$CO and C$^{18}$O emission.% and a currently incomplete understanding of the disk gas mass estimate from the $^{13}$CO and C$^{18}$O emission. 
The 1$\sigma$ dispersion in the distribution of the disk gas and dust masses is estimated to be 0.9 dex. 
We find that with such a large scatter, even if the disk gas masses are correlated with the disk dust masses, it is still expected that there is no correlation between the disk gas mass and the mass accretion rate seen in this sample. 
With the current uncertainties in deriving the disk gas mass, deeper observations, which can detect molecular lines in disks with a gas mass lower than 10$^{-5}$ $M_\sun$ and expand the range of disk masses in the sample, are needed to investigate the correlation between the disk gas mass and the mass accretion rate. 
}

\end{enumerate}

\begin{acknowledgements} 
This paper makes use of the following ALMA data: ADS/JAO.ALMA\#2013.1.00220.S. ALMA is a partnership of ESO (representing its member states), NSF (USA) and NINS (Japan), together with NRC (Canada), MOST and ASIAA (Taiwan), and KASI (Republic of Korea), in cooperation with the Republic of Chile. The Joint ALMA Observatory is operated by ESO, AUI/NRAO and NAOJ. We thank all the ALMA staff supporting this work. 
HWY, CFM, AM acknowledge ESO fellowship.
PMK acknowledges support from the Ministry of Science and Technology through grant MOST 104-2119-M-001-019-MY3 and from an Academia Sinica Career Development Award.
This work was partly supported by by the Italian Ministero dell'Istruzione, Universit\`a e Ricerca through the grant Progetti Premiali 2012 iALMA (CUP C52I13000140001), and by the Deutsche Forschungs-gemeinschaft (DFG, German Research Foundation) - Ref no. FOR 2634/1 TE 1024/1-1.
Astrochemistry in Leiden is supported by the Netherlands Research School for Astronomy (NOVA), by a Royal Netherlands Academy of Arts and Sciences (KNAW) professor prize, and by the European Union A-ERC grant 291141 CHEMPLAN.
\end{acknowledgements}

\begin{appendix}
\section{Distributions of S/N ratios in parameter space}\label{covar}
\begin{figure*}
\centering
\includegraphics[width=18cm]{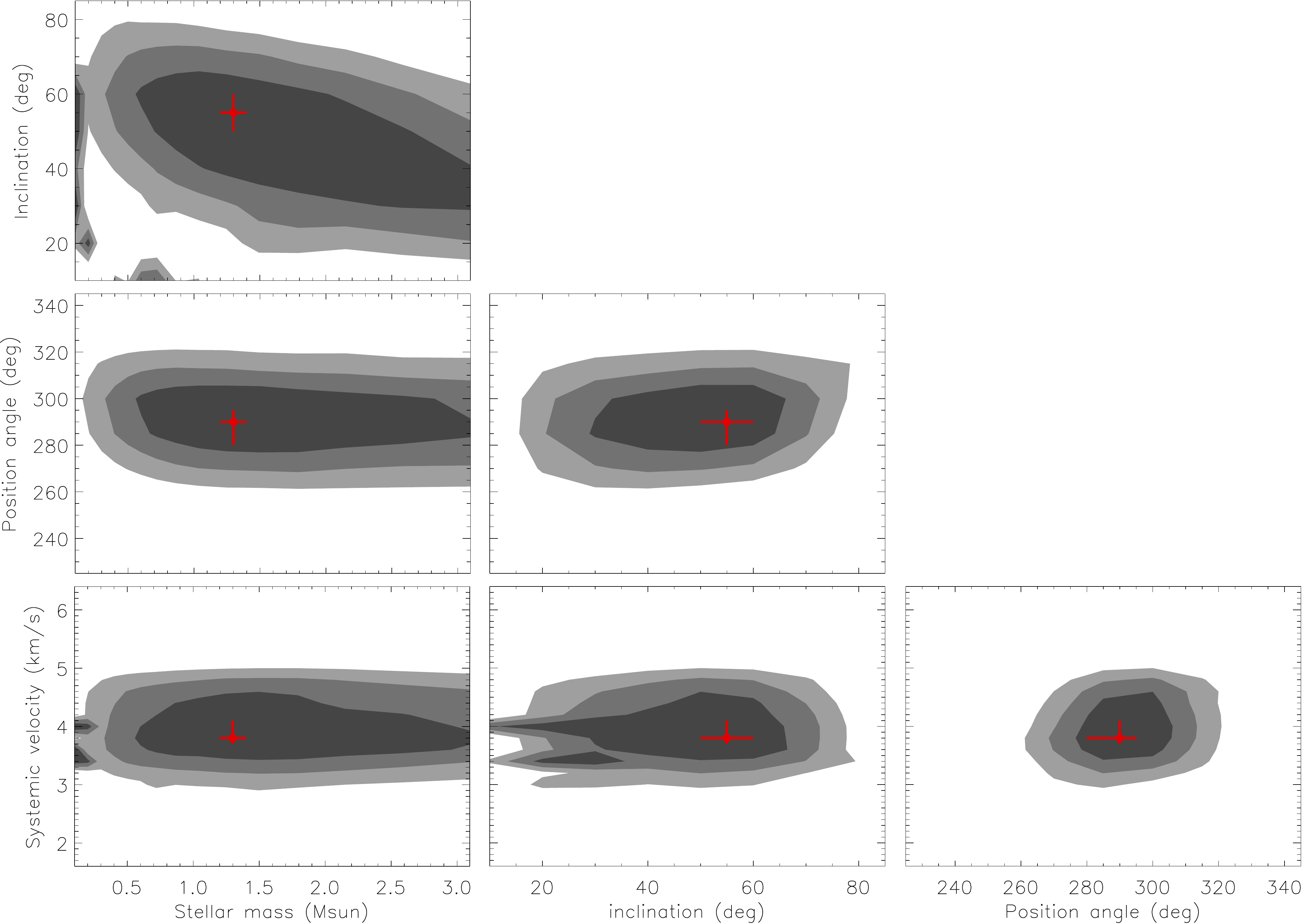}
\caption{Highest S/N ratios of the weighted integrated intensity of the aligned stacked spectra achieved with a given pair of the parameters by varying the other two parameters for \object{RY~Lup}. Contours from higher (dark) to lower (light) levels delineate the parameter space with achieved S/N ratios of the maximum S/N ratio $-1$, $-2$, and $-3$. Red data points with error bars show our measurements and uncertainties reported in Table \ref{test1}. The uncertainties are smaller than the parameter ranges delineated by the maximum S/N ratio $-1$ contour because of the additional constraints from the line profiles (see Section \ref{method}).}\label{covarfig1}
\end{figure*}

\begin{figure*}
\centering
\includegraphics[width=18cm]{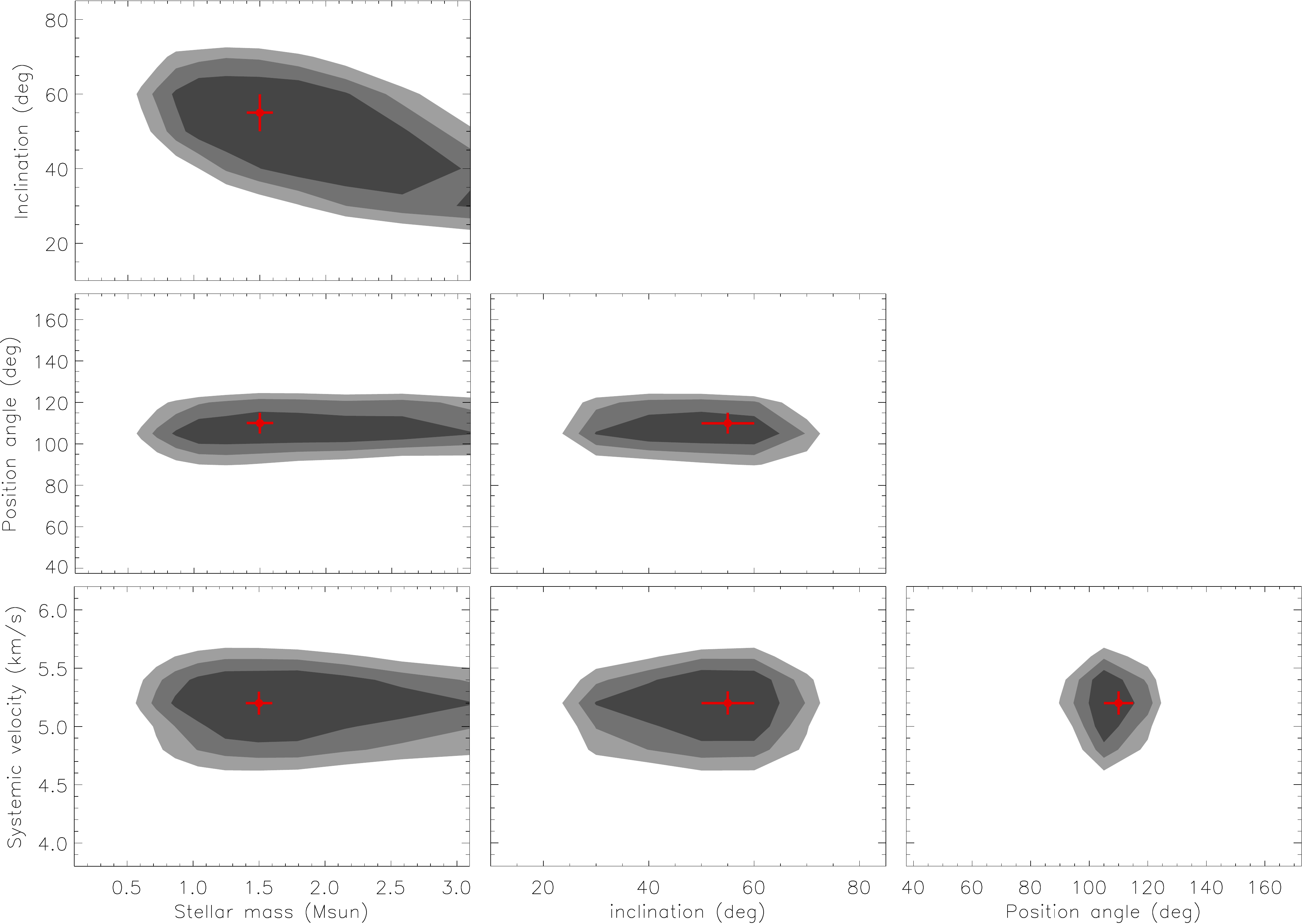}
\caption{Same as Fig.~\ref{covarfig1} but for \object{J16083070$-$3828268}.}\label{covarfig2}
\end{figure*}

\begin{figure*}
\centering
\includegraphics[width=18cm]{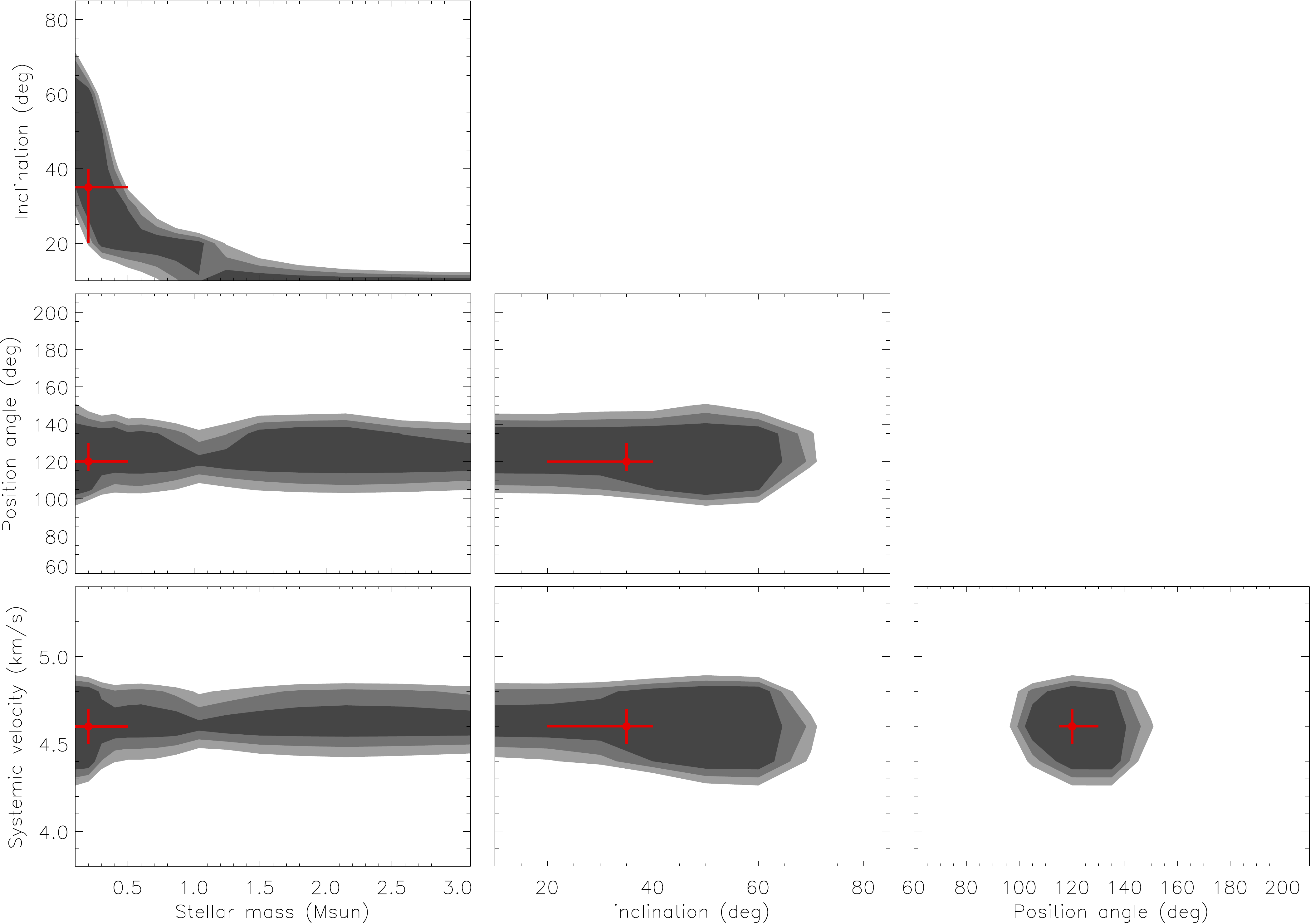}
\caption{Same as Fig.~\ref{covarfig1} but for \object{Sz~83}.}\label{covarfig3}
\end{figure*}

Figure \ref{covarfig1}--\ref{covarfig3} present the distributions of the highest S/N ratios of the weighted integrated intensity of the velocity aligned stacked spectra achieved with a given pair of the parameters by varying the other two parameters. 
The results of the analysis of the three Lupus sources, \object{RY~Lup}, \object{J16083070$-$3828268}, and \object{Sz~83} are presented. 
Red data points with error bars denote our measurements and the uncertainties.
Note that in our analysis, when estimating the uncertainties of our measurements, the information of the line profiles is also taken into account (Section \ref{method}).
These plots here only show the S/N ratios of the integrated intensity. 
Hence, the estimated uncertainties can be smaller than the parameter ranges delineated by the contour of the maximum S/N ratio $- 1$ of the integrated intensity.

In the case of \object{RY~Lup}, there are isolated regions showing high S/N ratios, for example, when the parameter $M_\star$ is lower than 0.5 $M_\sun$.  
The presence of these regions is caused by shifting high-noise channels to $V_{\rm sys}$ with these particular sets of parameters when aligning the spectra, 
and they are false signals. 
Thus, such isolated regions showing high S/N ratios are trimmed in our analysis. 

The distributions of the S/N ratios also show a clear correlation between the parameters $i$ and $M_\star$, where higher inclination angles tend to be associated with lower stellar masses. 
Thus, the uncertainty in $i$ can propagate to $M_\star$.
This is clearly seen in the case of \object{Sz~83}. 
The disk around \object{Sz~83} is close to face on, and thus its inclination angle is poorly constrained (Section \ref{demoob}).
That leads to a larger uncertainty in the estimated stellar mass.
On the other hand, there is no correlation found between other pairs in the parameter set.
Thus, they do not have any significant covariance, even for the pair of $i$ and PA.
This is different from estimating $i$ and PA from the intensity distributions of the continuum or molecular lines, 
where $i$ and PA can have a significant covariance. 
In particular, when the inclination is close to face on, the disk orientation cannot be constrained from the intensity distributions. 
In our analysis, the disk orientation PA is constrained from the direction of the velocity gradient due to disk rotation, and thus, it does not have a significant covariance with the inclination $i$.

\section{Simple disk models}\label{mkmod}
The model disks were made with two intensity distributions, uniform and Gaussian distributions, two stellar masses, 0.2 and 0.5 $M_\sun$, two inclination angles, 30\degr\ and 60\degr, and three different total integrated fluxes, 400, 600, and 800 mJy km s$^{-1}$. 
The position angles of their disk major axes were all fixed at 135\degr. 
For those uniform disks, the disk radii were adopted to be 1\arcsec, 
and for those Gaussian disks, the FWHM widths were set to be 1\arcsec. 
$V_{\rm sys}$ was adopted to be $V_{\rm LSR}$ of 3.7 km s$^{-1}$, the mean $V_{\rm sys}$ of Lupus YSOs \citep{Ansdell16}.
The distance was assumed to be 150 pc, the same as that to the Lupus I, II, and IV regions.
We first used the {\it Mirad} tasks, {\it imgen} and {\it velmodel}, to generate intensity distributions and velocity patterns of model disks, respectively. 
From those, model image cubes were produced using {\it Mirad} task, {\it velimage}, assuming a constant Gaussian 1$\sigma$ line width of 0.2 km s$^{-1}$. 
Then, we used CASA to simulate ALMA observations with the array configuration of C34-5 and a one-minute integration time on our model image cubes, resulting in an angular resolution of 0\farcs35 $\times$ 0\farcs29 and a noise level of 72 Jy Beam$^{-1}$ per channel, the same as the observations. 

\section{Detailed results of demonstration with models}\label{detail}
\begin{figure}
\centering
\includegraphics[width=8cm]{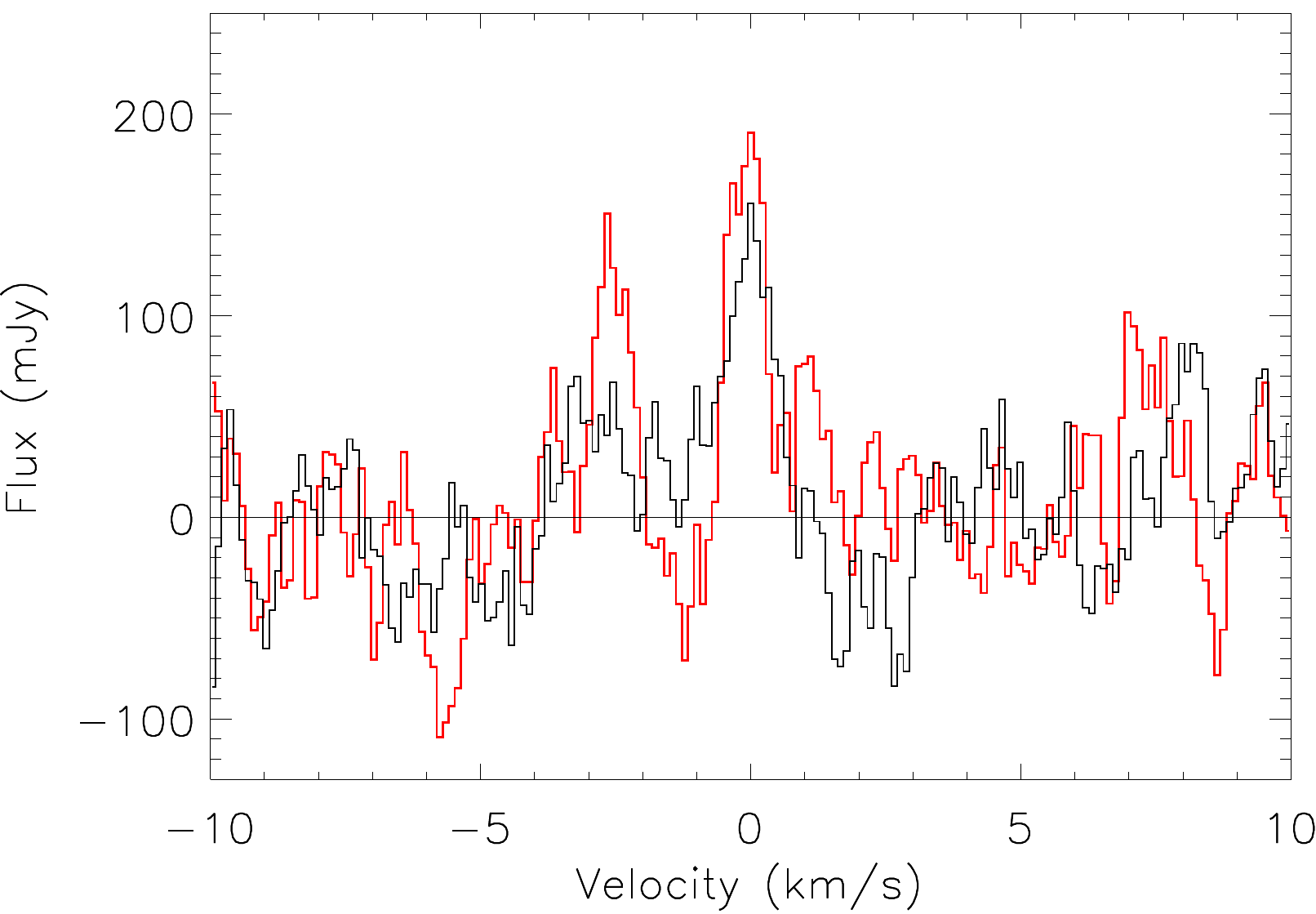}
\caption{Aligned stacked spectra of model 22 in Table \ref{test2}. Black and red histograms present the stacked spectra aligned with the input and measured parameters, respectively.}\label{model22}
\end{figure}

We applied our method on the synthetic images of the model disks to measure their $M_\star$, PA, $i$, and $V_{\rm sys}$. 
All the results are listed in Table \ref{test2}.
For the detected model disks with the integrated fluxes of 600 and 800 mJy km s$^{-1}$, 
the measured $M_\star$, PA, and $V_{\rm sys}$ are all consistent with the model inputs within the 1$\sigma$ uncertainties, except for model 24 and 11, where the measured $M_\star$ and PA are within the 2$\sigma$ uncertainties, respectively. 
%The difference between the measured and input $M_\star$, PA, and $V_{\rm sys}$ is less than 0.2 $M_\sun$, 15\degr, and 0.3 km s$^{-1}$, respectively. 
The inclination is more difficult to constrain because it requires more spatial information along the direction of the minor axis. 
In 9 out of 14 detected model disks with the integrated fluxes larger than 600 mJy km s$^{-1}$, 
the measured and input $i$ are consistent within the 1$\sigma$ uncertainties, 
and all of them are within the 3$\sigma$ uncertainties, except for model 23, where the measured $i$ significantly deviates from the input. 
Three of the model disks with the integrated fluxes of 400 mJy km s$^{-1}$ were detected with our method (model 4, 7, and 22). 
Even though their integrated fluxes are low, 
all the measured parameters in model 4 are consistent with the inputs within the 1$\sigma$ uncertainties,  
and in model 7, only the measured PA significantly deviates from the inputs, while the remaining parameters are all consistent with the inputs within the 1$\sigma$ uncertainties. 
The only exception among these detected model disks is model 22. 
For this model, all the measured parameters are different from the inputs. 
In Fig.~\ref{model22}, we compare the stacked spectra of model 22 aligned with the measured and input parameters.  
The one aligned with the measured parameters has a higher S/N ratio, 
which could be caused by its particular distribution of noise by coincidence. 
We have verified this by performing the same analysis on model 22 with different seeds of random noise, 
and there are cases with the measured parameters consistent with the inputs within the 1$\sigma$ uncertainties. 
Although in Fig.~\ref{model22} the parameters adopted to align the data are different, 
their measured fluxes are consistent within the uncertainties because our method conserves the flux \citep{Yen16}. 

\begin{table*}
\caption{Results of test using models}\label{test2}
\centering
\begin{tabular}{ccccccccccccc}
\hline\hline
 & \multicolumn{6}{c}{Input} && \multicolumn{5}{c}{Measurement} \\
 \cline{2-7} \cline{9-13}
Model & $I(r)$ & $M_\star$ & PA & i & $V_{\rm sys}$ & Flux && $M_\star$ & PA & i & $V_{\rm sys}$ & Flux \\
& & ($M_\sun$) & (\degr) & (\degr) & (km s$^{-1}$) & (mJy km s$^{-1}$) && ($M_\sun$) & (\degr) & (\degr) & (km s$^{-1}$) & (mJy km s$^{-1}$)\\
\hline \\
 1 & G & 0.2 & 135 & 30 & 3.7 & 400 &  & ... & ... & ... & ... & ... \smallskip\\
 2 & G & 0.2 & 135 & 30 & 3.7 & 600 &  & ... & ... & ... & ... & ... \smallskip\\
 3 & G & 0.2 & 135 & 30 & 3.7 & 800 &  & 0.2$\pm$0.1 & 140$\pm$5 &  35$\pm$5 & 3.7$\pm$0.1 &  532$\pm$45 \smallskip\\
 4 & G & 0.2 & 135 & 60 & 3.7 & 400 &  & 0.2$\pm$0.1 & 130$^{+10}_{- 5}$ & 60$^{+ 5}_{-15}$ & 3.8$^{+0.1}_{-0.2}$ &  297$\pm$37 \smallskip\\
 5 & G & 0.2 & 135 & 60 & 3.7 & 600 &  & 0.2$\pm$0.1 & 145$^{+ 5}_{-10}$ & 60$^{+ 5}_{-15}$ & 3.7$^{+0.1}_{-0.2}$ &  461$\pm$41 \smallskip\\
 6 & G & 0.2 & 135 & 60 & 3.7 & 800 &  & 0.2$^{+0.2}_{-0.1}$ & 130$^{+20}_{-10}$ & 45$^{+10}_{-15}$ & 3.9$^{+0.1}_{-0.3}$ &  402$\pm$47 \smallskip\\
 7 & G & 0.5 & 135 & 30 & 3.7 & 400 &  & 0.2$^{+0.4}_{-0.1}$ & 175$\pm$10 &  50$\pm$5 & 3.4$^{+0.5}_{-0.1}$ &  573$\pm$58 \smallskip\\
 8 & G & 0.5 & 135 & 30 & 3.7 & 600 &  & 0.3$^{+0.2}_{-0.1}$ & 135$\pm$10 & 45$^{+10}_{-15}$ & 3.7$^{+0.2}_{-0.1}$ &  432$\pm$46 \smallskip\\
 9 & G & 0.5 & 135 & 30 & 3.7 & 800 &  & 0.7$^{+1.3}_{-0.3}$ & 140$^{+ 5}_{-10}$ &  25$\pm$10 & 3.7$\pm$0.1 &  680$\pm$72 \smallskip\\
 10 & G & 0.5 & 135 & 60 & 3.7 & 400 &  & ... & ... & ... & ... & ... \smallskip\\
11 & G & 0.5 & 135 & 60 & 3.7 & 600 &  & 0.4$^{+0.3}_{-0.1}$ & 150$^{+ 5}_{-10}$ & 65$^{+ 5}_{-15}$ & 3.8$^{+0.3}_{-0.1}$ &  255$\pm$31 \smallskip\\
12 & G & 0.5 & 135 & 60 & 3.7 & 800 &  & 0.6$^{+1.0}_{-0.1}$ & 130$^{+10}_{-20}$ & 50$^{+ 5}_{-20}$ & 3.6$^{+0.2}_{-0.6}$ &  580$\pm$38 \smallskip\\
13 & U & 0.2 & 135 & 30 & 3.7 & 400 &  & ... & ... & ... & ... & ... \smallskip\\
14 & U & 0.2 & 135 & 30 & 3.7 & 600 &  & 0.2$^{+0.3}_{-0.1}$ & 135$\pm$10 & 40$^{+ 5}_{-20}$ & 3.7$^{+0.2}_{-0.1}$ &  485$\pm$65 \smallskip\\
15 & U & 0.2 & 135 & 30 & 3.7 & 800 &  & 0.2$^{+0.4}_{-0.1}$ & 135$^{+20}_{-10}$ & 35$^{+ 5}_{-15}$ & 3.6$\pm$0.1 &  568$\pm$63 \smallskip\\
16 & U & 0.2 & 135 & 60 & 3.7 & 400 &  & ... & ... & ... & ... & ... \smallskip\\
17 & U & 0.2 & 135 & 60 & 3.7 & 600 &  & 0.2$\pm$0.1 & 145$\pm$5 &  65$\pm$5 & 3.5$^{+0.2}_{-0.1}$ &  297$\pm$37 \smallskip\\
18 & U & 0.2 & 135 & 60 & 3.7 & 800 &  & 0.3$\pm$0.1 & 130$\pm$5 & 45$^{+ 5}_{-10}$ & 3.7$\pm$0.1 &  591$\pm$61 \smallskip\\
19 & U & 0.5 & 135 & 30 & 3.7 & 400 &  & ... & ... & ... & ... & ... \smallskip\\
20 & U & 0.5 & 135 & 30 & 3.7 & 600 &  & ... & ... & ... & ... & ... \smallskip\\
21 & U & 0.5 & 135 & 30 & 3.7 & 800 &  & 0.5$^{+1.5}_{-0.2}$ & 145$^{+ 5}_{-10}$ & 30$^{+10}_{-15}$ & 3.7$\pm$0.1 &  506$\pm$55 \smallskip\\
22 & U & 0.5 & 135 & 60 & 3.7 & 400 &  & 1.1$^{+0.4}_{-0.1}$ & 180$\pm$5 & 45$^{+ 5}_{-10}$ & 3.3$\pm$0.1 &  477$\pm$55 \smallskip\\
23 & U & 0.5 & 135 & 60 & 3.7 & 600 &  & 0.6$^{+1.2}_{-0.1}$ & 140$^{+20}_{- 5}$ & 35$^{+ 5}_{-15}$ & 3.7$^{+0.1}_{-0.2}$ & 1017$\pm$114 \smallskip\\
24 & U & 0.5 & 135 & 60 & 3.7 & 800 &  & 0.7$^{+0.2}_{-0.1}$ & 130$^{+10}_{- 5}$ &  50$\pm$5 & 3.6$^{+0.1}_{-0.2}$ &  831$\pm$81 \smallskip\\
%25 & G & 1.5 & 135 & 30 & 3.7 & 600 & & ... & ... & ... & ... & ... \smallskip\\
%26 & G & 1.5 & 135 & 60 & 3.7 & 600 & & 1.5$^{+1.0}_{-0.1}$ & 125$^{+15}_{-30}$ & 60$^{+5}_{-10}$ & 3.6$^+{1.2}_{0.1}$ \smallskip\\
%25 & G & 0.5 & 135 & 15 & 3.7 & 600 & & ... & ... & ... & ... & ... \smallskip\\
%25 & G & 0.5 & 135 & 15 & 3.7 & 800 & & 0.1$^{+0.8}_{-0}$ & 140$^{+15}$_{-15}$ & 30$^{+5}_{-20}$ & 3.6$^{+0.1}_{-0.2}$ \smallskip\\
\hline
\end{tabular}
\tablefoot{$I(r)$ is the intensity distribution of the model disk, which is a Gaussian (G) or uniform (U) distribution.}
%\tablebib{(1) \citet{branch83}; (2) \citet{phillips87}; (3) \citet{barbon90}}
\end{table*}

\section{Aligned stacked spectra of the full sample}\label{allspec}
\begin{figure*}
\centering
\includegraphics[width=17.5cm]{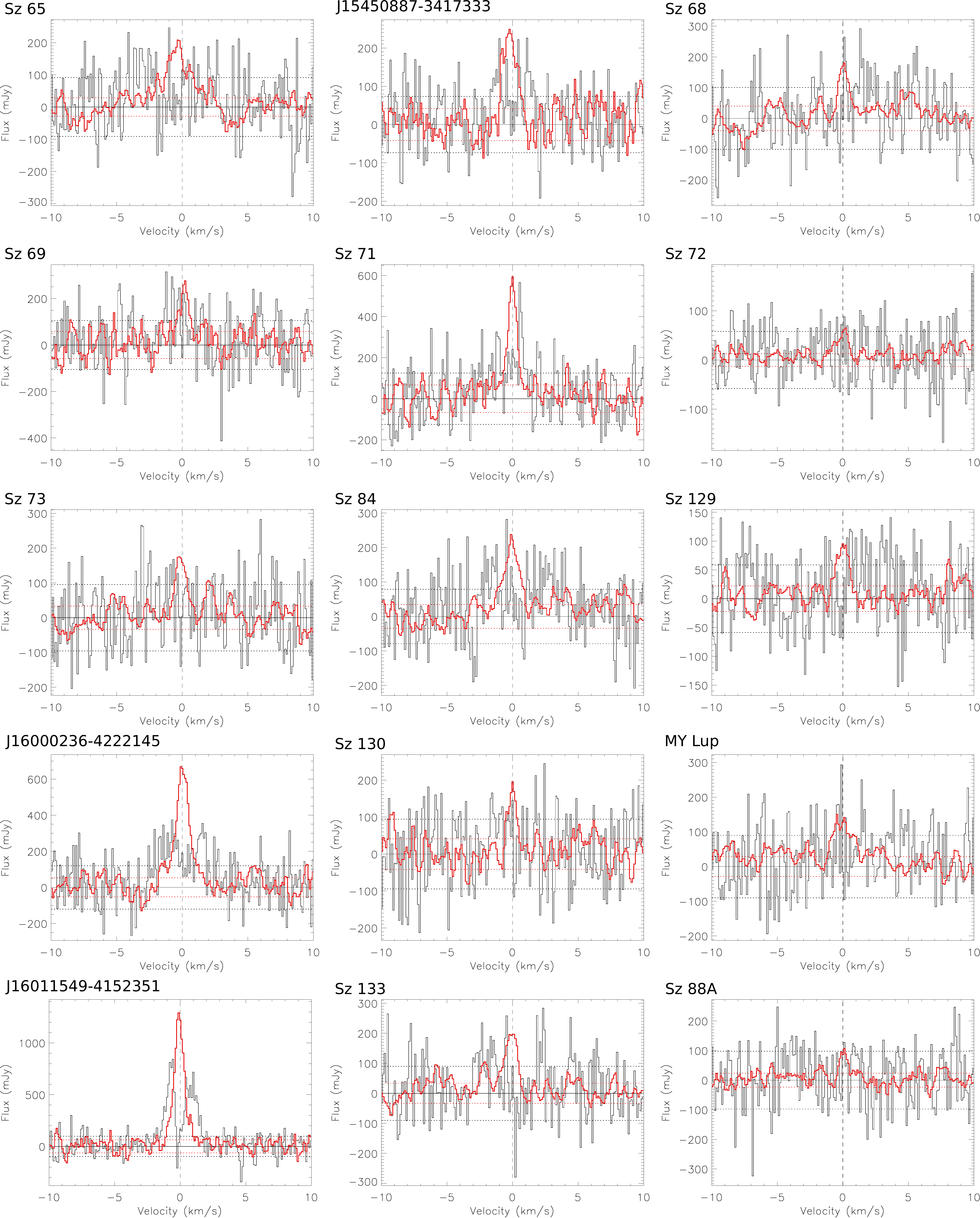}
\caption{Stacked $^{13}$CO spectra with (red) and without (black) alignment integrated over the disk area. Zero velocity refers to the measured $V_{\rm sys}$. Red and black horizontal dotted lines denote $\pm$1$\sigma$ levels with and without alignment, respectively.}\label{spec1}
\end{figure*}

Figure \ref{spec1} to \ref{spec5} present the velocity-aligned stacked spectra of the $^{13}$CO and C$^{18}$O lines of all the detected disks in comparison with the original non-aligned stacked spectra.

\begin{figure*}
\centering
\includegraphics[width=17.5cm]{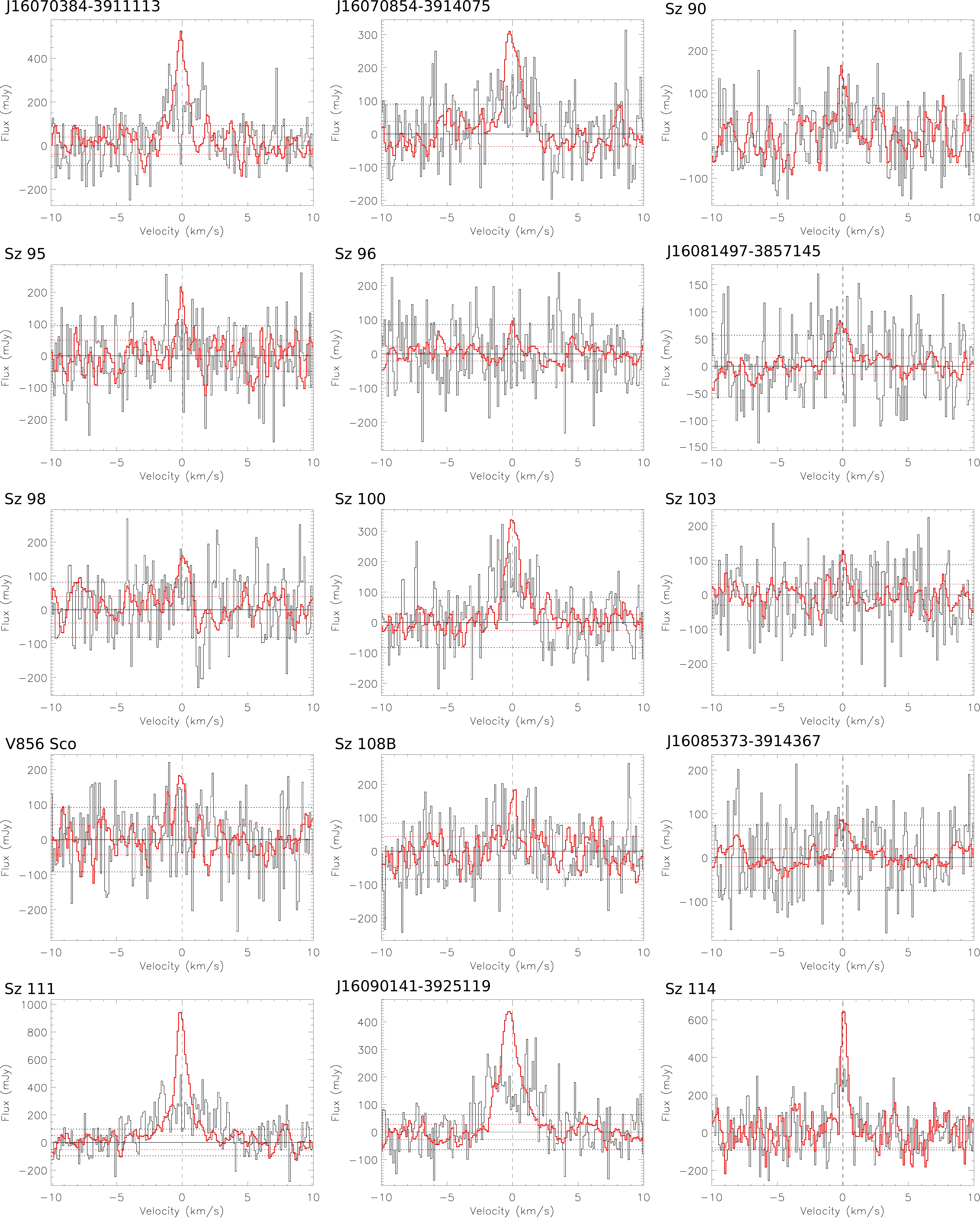}
\caption{Same as Fig.~\ref{spec1}.}
\end{figure*}

\begin{figure*}
\centering
\includegraphics[width=17.5cm]{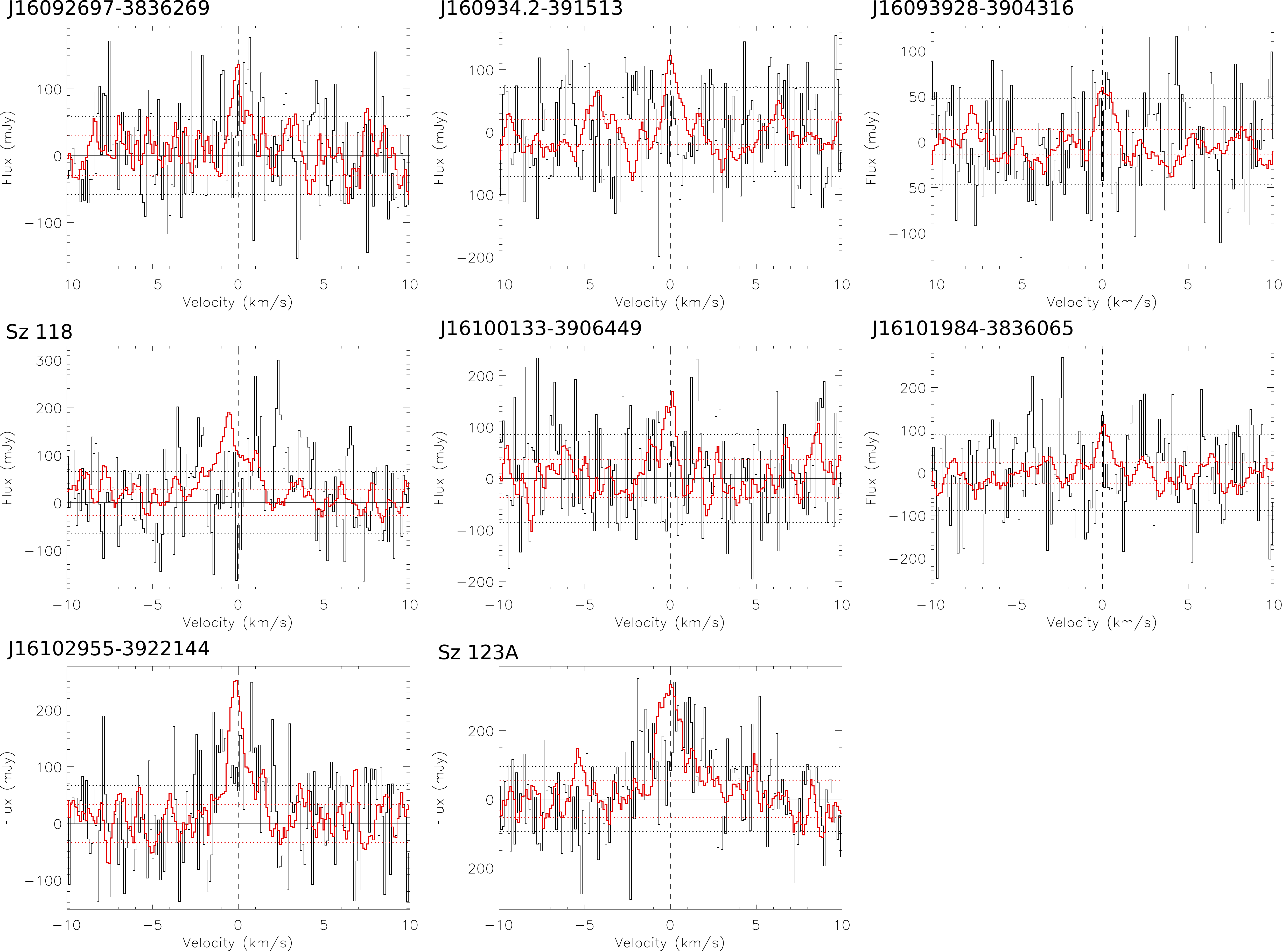}
\caption{Same as Fig.~\ref{spec1}.}
\end{figure*}

\begin{figure*}
\centering
\includegraphics[width=17.5cm]{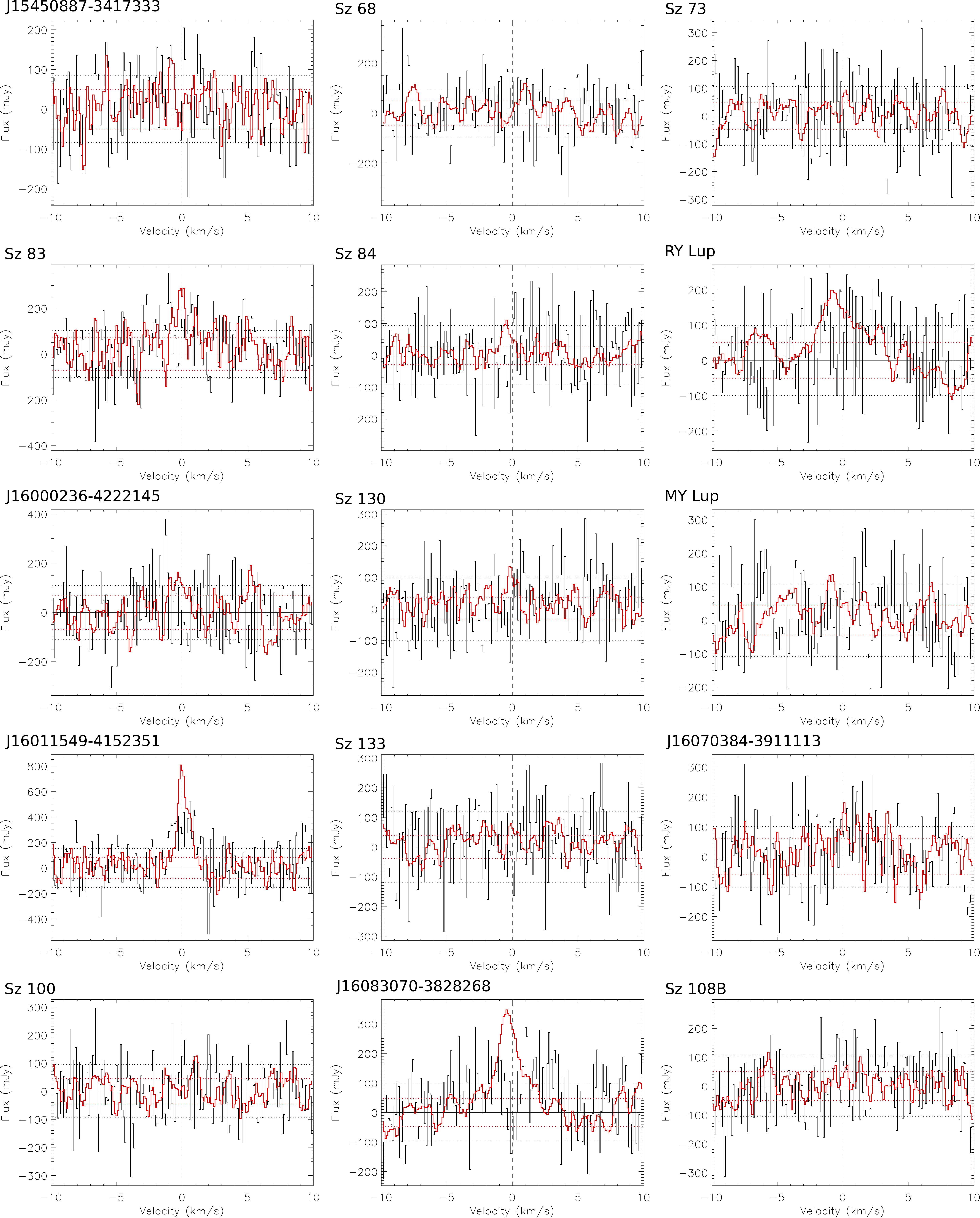}
\caption{Same as Fig.~\ref{spec1} but for the C$^{18}$O emission.}\label{spec2}
\end{figure*}

\begin{figure*}
\centering
\includegraphics[width=17.5cm]{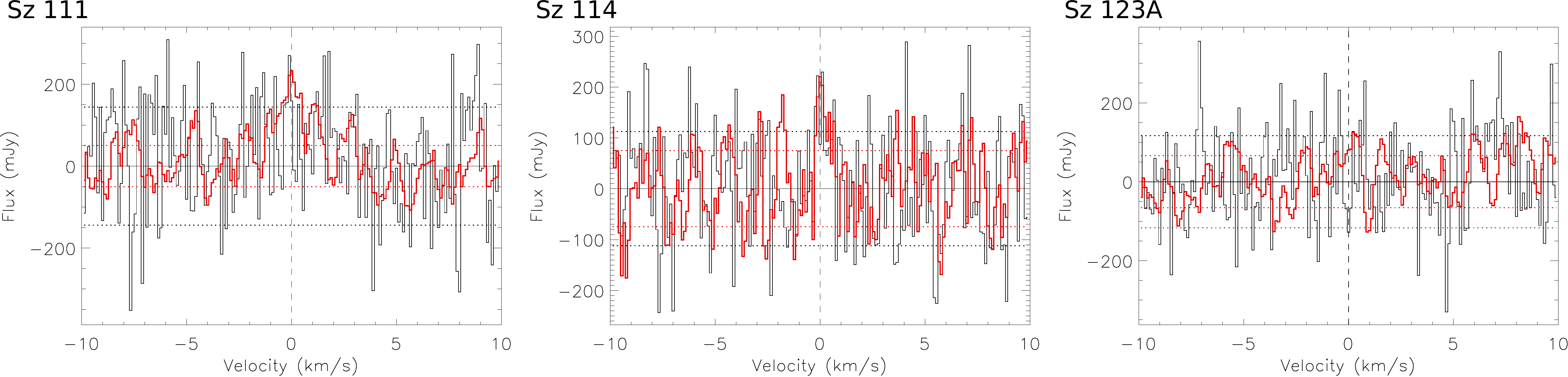}
\caption{Same as Fig.~\ref{spec2}.}\label{spec5}
\end{figure*}

\end{appendix}

\end{document}